# Fourier modal method and coordinate transformation method under nonclassical electromagnetic boundary condition for the electromagnetism of mesoscale metallic nanostructures

Haitao Liu[1,2,*]

[1]*Institute of Modern Optics, College of Electronic Information and Optical Engineering, Nankai University, Tianjin, 300350, China*
[2]*Tianjin Key Laboratory of Micro-scale Optical Information Science and Technology, Tianjin, 300350, China*

The optical response of mesoscale metallic nanostructures (MMNSs) with feature sizes down to extreme nanometer scales is largely affected by the nonclassical quantum effects, which can be comprehensively described by the nonclassical electromagnetic boundary condition (NEBC) incorporating surface-response Feibelman $d$-parameters. Here we report the Fourier modal method (FMM) and the coordinate transformation method (C method) under the NEBC, which are built up by incorporating the NEBC into a recently reported 3D-C method [Opt. Express 29, 1516 (2021)] that is applicable to the general three-dimensional (3D) photonic structures with curved boundaries. The validity and accuracy of the proposed method are confirmed numerically through a comparison with other full-wave method incorporating the NEBC. The present work marries the NEBC and the well-developed modal methods of FMM and C method, thus bringing the advantages of these modal methods in physical intuitiveness and computational efficiency to the electromagnetic modeling of nonclassical quantum effects in the MMNSs.

## I. INTRODUCTION

Mesoscale metallic nanostructures (MMNSs) with feature sizes (such as the size of nanogap) down to extreme nanometer scales can achieve an extreme nanoscale confinement of plasmonic electromagnetic field and thus drastically enhance the light-matter interactions [1-5]. The optical response of MMNSs is largely affected by the nonclassical quantum effects [6-11] and cannot be faithfully described by the classical Maxwell's equations [9-14]. The first-principle based time-dependent density functional theory can rigorously describe the nonclassical quantum effects but at the expense of a huge computational amount [15-17]. Alternatively, by modifying the classical electromagnetic boundary condition (CEBC) at the metal-dielectric interface to be nonclassical electromagnetic boundary condition (NEBC) [11,18] which incorporates two first-principle based surface-response Feibelman $d$-parameters [6,19,20], $d_\perp$ and $d_\parallel$, the nonclassical quantum effects including the nonlocality, surface Landau damping and electron spill-in/out [6,9,10,21] can be comprehensively described with a first-principle level accuracy. Besides, in the bulk region away from the interface of NEBC, classical Maxwell's equations are still satisfied by the electromagnetic field [11,18,21], which is beneficial for incorporating the NEBC into the various well-developed methods for solving the classical Maxwell's equations [22-25].

The development of full-wave numerical methods for solving Maxwell's equations under the NEBC is still in its early stage. In this aspect, the boundary element method incorporating the NEBC with local [26] or nonlocal [27] Feibelman $d$-parameters has been reported. By introducing a scalar auxiliary potential to ensure the numerical stability, the NEBC is incorporated into the finite element method (FEM) which is carried out with the commercial COMSOL Multiphysics software [11]. Semianalytical quasinormal mode (QNM) expansion theories are also developed for solving the source-excited electromagnetic field under the NEBC as an expansion upon the basis of nonclassical QNMs under NEBC [28] or upon the basis of classical QNMs under CEBC with an analytical treatment of the NEBC [29].

The Fourier modal method (FMM) [25,30,31], also known as the rigorous coupled-wave analysis [32], is one of the most popular methods for modelling periodic photonic structures. As a modal method, the FMM [25,33-37] expresses the electromagnetic field as a superposition of $z$-propagating waveguide modes [38,39] with an analytical $z$-coordinate dependence (in each $z$-invariant layer that constitutes the whole structure), and can obtain the full scattering matrix [40] of the structure by solving the Maxwell's equations only once, thus possessing rich physical intuitiveness and a high computational efficiency compared with other full-wave numerical methods [23,24]. Besides, the electromagnetic field is discretized into Fourier series (along the transversal $x$- and $y$-directions) in the FMM, which may benefit from the well-established theories of Fourier analysis [41]. The FMM has experienced substantial developments such as the correct Fourier factorization rules [33-37] and adaptive spatial resolution (ASR) [42,43] for improving the convergence, the scattering-matrix algorithms for improving the numerical stability [40], the perfectly matched layers (PMLs) [44] for aperiodic structures, and the matched coordinate transformation [45,46] to avoid the zigzag approximation of curved boundaries in $x$- and $y$-directions.

For the structures with curved boundaries between adjacent *z*-invariant layers, classical FMM requires a staircase approximation of the curved boundaries [47,48], which can be avoided by a coordinate transformation method (C method) [49-57] where a curvilinear coordinate system is adopted to map the curved boundaries to planar boundaries, and then the Maxwell's equations under the curvilinear coordinate system are solved by using the algorithm of the classical FMM. To improve the limited applicability of the C method only to two-dimensional structures (i.e. invariant along a transversal *y*-direction) [49,52-56] or structures with identical-profile curved boundaries between adjacent *z*-invariant layers [50,51], a 3D-C method applicable to the general case of three-dimensional (3D, i.e. varying along both transversal *x*- and *y*-directions) structures with different-profile curved boundaries between adjacent *z*-invariant layers is proposed [57]. However, despite of the above tremendous developments, the FMM and the C method under the NEBC have not been built up.

In this paper, the 3D-C method (along with the FMM as the special case of the 3D-C method) under the NEBC is reported, so as to bring the advantages of these modal methods in physical intuitiveness and computational efficiency to the electromagnetic modeling of nonclassical quantum effects in the MMNSs. Here the 3D-C method and the FMM are collectively referred to as coordinate-transformation Fourier modal method (CFMM) for brevity. The establishment of the CFMM under NEBC relies on several key steps. First, the tangential electric-field and magnetic-field NEBCs under the curvilinear coordinate system of CFMM (abbreviated as CFMM-NEBCs) are derived (Secs. IIIA and IIIB). Second, by properly reformulating the derived CFMM-NEBCs such that the correct Fourier factorization rules [35] are applicable, the Fourier representations of the CFMM-NEBCs are obtained (Secs. IIIC and IIID). Last, the Fourier representations of the CFMM-NEBCs are incorporated into the scattering-matrix algorithm [57], so that the unknown coefficients of waveguide modes (and resultantly, the electromagnetic field) can be finally obtained (Secs. IIIE). The validity and accuracy of the proposed CFMM under NEBC are tested numerically in comparison with the full-wave FEM incorporating the NEBC [11] (Sec. IV).

This paper is organized as follows. In Sec. II, a brief review of the 3D-C method under CEBC is provided. In Sec. III, the CFMM under NEBC is built up. In Sec. IV, a numerical test of the CFMM under NEBC is provided. The conclusions are summarized in Sec. V.

## II. A BRIEF REVIEW OF THE 3D-C METHOD UNDER CEBC

As a preparation for building up the CFMM under NEBC, in this section we will provide a brief review of the 3D-C method under CEBC proposed in Ref. [57], which extends the C method [49-56] to the general case of 3D structures with different-profile curved boundaries between adjacent *z*-invariant layers. All the symbols in this section follow those in Ref. [57] for a convenient reading.

As sketched in Fig. 1, in the 3D-C method under CEBC, the whole structure is decomposed into *L* central layers along with a bottom and a top semi-infinite layers (called layer 0, 1, …, *L*+1 from bottom to top). The medium is assumed to be generally anisotropic medium described by a permittivity tensor **ε**(**r**) and a permeability tensor **μ**(**r**) as functions of the Cartesian spatial coordinate **r**=(*x,y,z*). The bottom boundary of layer *l* (*l*=1, 2, …, *L*+1) is assumed to be a generally curved surface $z=f^{(l)}(x,y)$, and can be of generally different profile for different *l*, i.e. $f^{(l)}(x,y)-f^{(m)}(x,y)$ (with $l \neq m$) may vary with *x* and *y* in general. Each layer [or, **ε**(**r**) and **μ**(**r**)] is assumed to be invariant in the *z*-direction, and the whole structure [or, **ε**(**r**), **μ**(**r**) and $f^{(l)}(x,y)$] is assumed to be periodic in the *x*- and *y*-directions. For a structure that is aperiodic in *x*- or *y*-direction (see Fig. 1 for instance), it can be mapped to be an artificial periodic structure by using the PML [44].

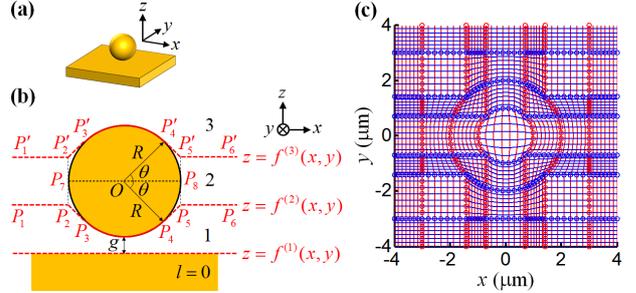

FIG. 1. **(a)** Structure of the numerical example of a single gold nanosphere on a flat gold substrate in water. **(b)** Definition of *z*-invariant layers *l*=0, 1, 2, 3 (*L*=2) for the structure in (a). The coordinate origin *O* is set at the center of the nanosphere. The boundary surfaces $z=f^{(l)}(x,y)$ (*l*=1, 2, 3) between adjacent layers are shown by the red lines. **(c)** Curves of $u=u_C$ (red) or $v=v_C$ (blue) for the matched coordinate transformation $x=x(u,v)$ and $y=y(u,v)$. There are $w_x=w_y=6\mu m$, $R=2\mu m$ and $\theta=60°$ for (c), and these values are different from the actual values of the numerical example for a clear illustration.

For layers *l*−1 and *l* separated by $z=f^{(l)}(x,y)$, the 3D-C method adopts a curvilinear coordinate system $(u^1,u^2,u^3)=(u,v,w)$ defined as [57],

$$x = x(u,v), \; y = y(u,v), \; z = w + f(x(u,v), y(u,v)), \quad (1)$$

where $x=x(u,v)$ and $y=y(u,v)$ represent the matched coordinate transformation [45,46] to be explained hereafter, and $f(x,y)=f^{(l)}(x,y)$ with the superscript neglected for simplicity. If $f^{(l)}(x,y)$ is a constant for any *l*=1, 2, …, *L*+1, then the 3D-C method reduces to the classical FMM [25]. In this paper, the 3D-C method and its special case of FMM are collectively referred to as CFMM for brevity.

Under (*u,v,w*), one can solve the covariant form of the Maxwell's equations [57],

$$\begin{vmatrix} \mathbf{e}_1 & \mathbf{e}_2 & \mathbf{e}_3 \\ \dfrac{\partial}{\partial u^1} & \dfrac{\partial}{\partial u^2} & \dfrac{\partial}{\partial u^3} \\ E_1 & E_2 & E_3 \end{vmatrix} = i\omega(\mathbf{e}_1,\mathbf{e}_2,\mathbf{e}_3)V \begin{bmatrix} \mu^{1,1} & \mu^{1,2} & \mu^{1,3} \\ \mu^{2,1} & \mu^{2,2} & \mu^{2,3} \\ \mu^{3,1} & \mu^{3,2} & \mu^{3,3} \end{bmatrix}\begin{bmatrix} H_1 \\ H_2 \\ H_3 \end{bmatrix},$$

(2a)

$$\begin{vmatrix} \mathbf{e}_1 & \mathbf{e}_2 & \mathbf{e}_3 \\ \dfrac{\partial}{\partial u^1} & \dfrac{\partial}{\partial u^2} & \dfrac{\partial}{\partial u^3} \\ H_1 & H_2 & H_3 \end{vmatrix} = -i\omega(\mathbf{e}_1,\mathbf{e}_2,\mathbf{e}_3)V \begin{bmatrix} \varepsilon^{1,1} & \varepsilon^{1,2} & \varepsilon^{1,3} \\ \varepsilon^{2,1} & \varepsilon^{2,2} & \varepsilon^{2,3} \\ \varepsilon^{3,1} & \varepsilon^{3,2} & \varepsilon^{3,3} \end{bmatrix}\begin{bmatrix} E_1 \\ E_2 \\ E_3 \end{bmatrix}.$$

(2b)

In Eq. (2), |·| means determinant, and the covariant basis vectors are given by $(\mathbf{e}_1,\mathbf{e}_2,\mathbf{e}_3)=(\mathbf{x},\mathbf{y},\mathbf{z})\mathbf{P}$, where $(\mathbf{x},\mathbf{y},\mathbf{z})$ are the Cartesian basis vectors, and the $i$-th row and $j$-th column element of matrix $\mathbf{P}$ is $(\mathbf{P})_{i,j}=\partial x_i/\partial u^j$ with $(x_1,x_2,x_3)=(x,y,z)$. The unknowns $E_i=\mathbf{e}_i\cdot\mathbf{E}$ and $H_i=\mathbf{e}_i\cdot\mathbf{H}$ ($i=1,2,3$) to be solved are the covariant components of the electric vector $\mathbf{E}$ and magnetic vector $\mathbf{H}$. $\omega$ is the angular frequency. $\varepsilon^{i,j}=\mathbf{e}^i\cdot\boldsymbol{\varepsilon}\cdot\mathbf{e}^j$ and $\mu^{i,j}=\mathbf{e}^i\cdot\boldsymbol{\mu}\cdot\mathbf{e}^j$ are the contravariant components of permittivity tensor $\boldsymbol{\varepsilon}$ and permeability tensor $\boldsymbol{\mu}$, respectively, and the contravariant basis vectors are given by $(\mathbf{e}^1,\mathbf{e}^2,\mathbf{e}^3)=(\mathbf{x},\mathbf{y},\mathbf{z})\mathbf{Q}$ with

$$\mathbf{Q} = (\mathbf{P}^{\mathrm{T}})^{-1} = \begin{bmatrix} \dfrac{y_v}{V} & -\dfrac{y_u}{V} & -f_x \\ -\dfrac{x_v}{V} & \dfrac{x_u}{V} & -f_y \\ 0 & 0 & 1 \end{bmatrix},$$

(3)

where $x_u$, $x_v$, $y_u$, $y_v$, $f_x$ and $f_y$ are defined in the same way as $x_u=\partial x/\partial u$, and $V=x_u y_v - x_v y_u$. It is remarkable that Eq. (2) has the same form as the Maxwell's equations under $(x,y,z)$ (with $V\varepsilon^{i,j}$ and $V\mu^{i,j}$ as the equivalent permittivity and permeability tensors), which is the well-known form invariance of Maxwell's equations under a curvilinear coordinate system [38,58,59], so that the algorithm of FMM [37] for solving the Maxwell's equations of anisotropic medium under $(x,y,z)$ can be applied as well for solving Eq. (2), as done in the 3D-C method [57].

Equation (2) is written using the international system of units. In this paper (also in Ref. [57]), we instead use the Gaussian system of units ($\varepsilon=\mu=1$ in the vacuum) for the convenience of numerical implementation, so that the $\omega$ in Eq. (2) actually becomes the wavenumber $k_0$ in the vacuum ($k_0=2\pi/\lambda$ with $\lambda$ being the wavelength).

Compared with solving the Maxwell's equations under $(x,y,z)$ by using the classical FMM [25], the advantages of solving Eq. (2) lie in the fact that the curved boundary $z=f(x,y)$ under $(x,y,z)$ is mapped to a flat boundary $w=0$ under $(u,v,w)$, so that the staircase approximation [47,48] of $z=f(x,y)$ adopted in the classical FMM under $(x,y,z)$ can be avoided. Besides, by defining the matched coordinate transformation [45,46] $x=x(u,v)$ and $y=y(u,v)$ such that $[x=x(u_C,v), y=y(u_C,v)]$ and $[x=x(u,v_C), y=y(u,v_C)]$ can cover the curves $C$ at the discontinuities of $V\varepsilon^{i,j}$ and $V\mu^{i,j}$ [or, of $\varepsilon(\mathbf{r})$, $\mu(\mathbf{r})$, $\partial f(x,y)/\partial x$ and $\partial f(x,y)/\partial y$], $C$ can be mapped to $u=u_C$ or $v=v_C$, so that the zigzag approximation [36] of $C$ adopted in the classical FMM under $(x,y,z)$ can be avoided by solving Eq. (2).

The PMLs [44] for aperiodic structures and the ASR [42,43] for enhancing the convergence can be incorporated by further introducing a coordinate transformation $u=G_1(u')$ and $v=G_2(v')$, which are maps of $R\to C$ with $u'$ and $v'$ in the region of PMLs, and are maps of $R\to R$ with $u'$ and $v'$ in the computational window truncated by the PMLs to perform the ASR. Then Eq. (2) with $(u',v',w)$ as the new differential variables should be solved after an replacement,

$$\dfrac{\partial}{\partial u}=g_1(u')\dfrac{\partial}{\partial u'},\quad \dfrac{\partial}{\partial v}=g_2(v')\dfrac{\partial}{\partial v'},\quad (4)$$

where $g_1(u')=[dG_1(u')/du']^{-1}$ and $g_2(v')=[dG_2(v')/dv']^{-1}$ are set to be continuous functions of $u'$ and $v'$, so that the Laurent's rule [35] can be applied for the Fourier factorization of the products in the right side of Eq. (4).

According to the *correct Fourier factorization rules* [35] including a Laurent's rule and an inverse rule, if either $f(u')$ or $g(u')$ is a continuous function, then the Fourier expansion coefficients of the product $h(u')=f(u')g(u')$ can be obtained from those of $f(u')$ and $g(u')$ by applying the Laurent's rule; if neither $f(u')$ nor $g(u')$ is a continuous function but $h(u')=f(u')g(u')$ is a continuous function, then the Fourier expansion coefficients of $h(u')$ can be obtained from those of $f(u')$ and $g(u')$ by appling the inverse rule. More detailed introduction about the correct Fourier factorization rules can be found in Supplemental Material Sec. S1A [60].

For the $z$-invariant layer $l$ ($l=0, 1, …, L+1$) under the curvilinear coordinate system of Eq. (1) defined by $z=f^{(l')}(x,y)$ ($l'=l$ and $l+1$ for the bottom and top boundaries of the central layer $l$, respectively, $l'=1$ and $L+1$ for the boundaries of the bottom and top semi-infinite layers, respectively), one can apply the correct Fourier factorization rules [35] to obtain the Fourier representation of the covariant-form Maxwell's equations (2) (Supplemental Material Sec. S1 [60]) [37], which then gives the modal solutions of up-going and down-going electromagnetic fields respectively expressed as [57],

$$\boldsymbol{\psi}_{l,+}^{(l')}(u',v',w^{(l')})=\sum_{r=1}^{N}u_{l,r}^{(l')}\boldsymbol{\psi}_{l,+,r}^{(l')}(u',v')\exp(ik_{l,+,r}^{(l')}w^{(l')}),\quad (5a)$$

$$\boldsymbol{\psi}_{l,-}^{(l')}(u',v',w^{(l')})=\sum_{r=1}^{N}d_{l,r}^{(l')}\boldsymbol{\psi}_{l,-,r}^{(l')}(u',v')\exp(ik_{l,-,r}^{(l')}w^{(l')}),\quad (5b)$$

where the subscript $l$ and superscript $l'$ are respectively used as the indices of layer and curvilinear coordinate system *throughout the paper*, $w^{(l')}=z-f^{(l')}(x,y)$, $\boldsymbol{\psi}=[E_1,E_2,E_3,H_1,H_2,H_3]^{\mathrm{T}}$ is a column vector containing all the covariant electromagnetic-field components, $\boldsymbol{\psi}_{l,+,r}^{(l')}$ and $\boldsymbol{\psi}_{l,-,r}^{(l')}$ represent the $r$-th up-going and down-going waveguide eigenmodes [38,39], respectively, $u_{l,r}^{(l')}$ and $d_{l,r}^{(l')}$ represent the corresponding mode coefficients, and $k_{l,+,r}^{(l')}$ and $k_{l,-,r}^{(l')}$ are the corresponding mode propagation constants [satisfying $\text{Re}(k_{l,+,r}^{(l')}) + \text{Im}(k_{l,+,r}^{(l')}) > 0$ and $\text{Re}(k_{l,-,r}^{(l')}) + \text{Im}(k_{l,-,r}^{(l')}) < 0$]. The $\psi_{l,\pm,r}^{(l')}(u',v')$ is expressed as a truncated Fourier series,

$$\psi_{l,\pm,r}^{(l')}(u',v') = \sum_{m=-M_x}^{M_x} \sum_{p=-M_y}^{M_y} F_{(u',v')}\{\psi_{l,\pm,r}^{(l')}\}_{(m,p)} \times \exp[i(k'_{u,m}u' + k'_{v,p}v')], \quad (6)$$

where an operator $F_{(u',v')}\{\cdot\}_{(m,p)}$ of calculating the Fourier expansion coefficient is defined as,

$$F_{(u',v')}\{\psi_{l,\pm,r}^{(l')}\}_{(m,p)} = \frac{1}{\Lambda'_u \Lambda'_v} \int_{u'_0}^{u'_0+\Lambda'_u} du' \int_{v'_0}^{v'_0+\Lambda'_v} dv' \psi_{l,\pm,r}^{(l')}(u',v') \times \exp[-i(k'_{u,m}u' + k'_{v,p}v')]. \quad (7)$$

In Eqs. (6) and (7), $k'_{u,m} = k'_{u,0} + m2\pi/\Lambda'_u$ and $k'_{v,p} = k'_{v,0} + p2\pi/\Lambda'_v$ with $m$ an $p$ being integers, $[u'_0, u'_0+\Lambda'_u]$ and $[v'_0, v'_0+\Lambda'_v]$ are one periods with respect to $u'$ and $v'$, and $k'_{u,0}$ and $k'_{v,0}$ are the pseudo-periodic phase-shift constants with respect to $u'$ and $v'$, respectively. There is $k'_{u,0}=0$ or $k'_{v,0}=0$ for the structure that is aperiodic in $x$- or $y$-direction, respectively [44]. The $F_{(u',v')}\{\psi_{l,\pm,r}^{(l')}\}_{(m,p)}$ (and resultantly, the $\psi_{l,\pm,r}^{(l')}$) and $k_{l,\pm,r}^{(l')}$ are obtained by solving a matrix eigenvalue problem derived from Eq. (2) with the algorithm of FMM (Supplemental Material Sec. S1C [60]), and the $u_{l,r}^{(l')}$ and $d_{l,r}^{(l')}$ are unknowns to be determined by matching the field-continuity boundary condition between adjacent $z$-invariant layers as explained in the following.

By adopting the hybrid-spectrum method proposed in Ref. [54] for 2D structures and extended in Ref. [57] to 3D structures, the electromagnetic field in the central layer $l$ ($l=1, 2, …, L$) can be expressed as

$$\Phi_l = \Phi_{l,+}^{(l)}(u',v',w^{(l)}) + \Phi_{l,-}^{(l+1)}(u',v',w^{(l+1)}), \quad (8)$$

where $\Phi = [\mathbf{E}, \mathbf{H}]^T$ is a column vector composed of the electric vector $\mathbf{E}$ and magnetic vector $\mathbf{H}$, $\Phi_l = [\mathbf{E}_l, \mathbf{H}_l]^T$, and $\Phi_{l,\pm}^{(l')}(u',v',w^{(l')}) = [\mathbf{E}_{l,\pm}^{(l')}, \mathbf{H}_{l,\pm}^{(l')}]$ ($l'=l, l+1$) with

$$\mathbf{E}_{l,\pm}^{(l')} = E_{1,l,\pm}^{(l')} \mathbf{e}^1 + E_{2,l,\pm}^{(l')} \mathbf{e}^2 + E_{3,l,\pm}^{(l')} \mathbf{e}^{3,(l')}, \quad (9a)$$

$$\mathbf{H}_{l,\pm}^{(l')} = H_{1,l,\pm}^{(l')} \mathbf{e}^1 + H_{2,l,\pm}^{(l')} \mathbf{e}^2 + H_{3,l,\pm}^{(l')} \mathbf{e}^{3,(l')}. \quad (9b)$$

In Eq. (9), the $E_{i,l,\pm}^{(l')}$ and $H_{i,l,\pm}^{(l')}$ ($i=1, 2, 3$) are simply the covariant field components of $\psi_{l,\pm}^{(l')}$ in Eq. (5), and ($\mathbf{e}^1, \mathbf{e}^2, \mathbf{e}^{3,(l')}$) are the ($\mathbf{e}^1, \mathbf{e}^2, \mathbf{e}^3$) defined in Eq. (3), where $\mathbf{e}^1$ and $\mathbf{e}^2$ are shared by all the layer boundaries, and $\mathbf{e}^{3,(l')}$ is defined by $z=f^{(l')}(x,y)$. One can see that in Eq. (8), the other two fields $\Phi_{l,+}^{(l+1)}$ and $\Phi_{l,-}^{(l)}$ in layer $l$ are never used. The electromagnetic field in the bottom ($l=0$) or top ($l=L+1$) semi-infinite layer is given by Eq. (8) with the first term replaced by $\Phi_{l,+}^{(l+1)}(u',v',w^{(l+1)})$ or the second term replaced by $\Phi_{l,-}^{(l)}(u',v',w^{(l)})$, respectively. Then the continuity boundary condition of tangential electromagnetic-field components at $z=f^{(l)}(x,y)$ ($l=1, 2, …, L+1$) can be expressed as,

$$\mathbf{e}_i^{(l)} \cdot \mathbf{E}_{l-1}\big|_{z=f^{(l)}(x,y)} = \mathbf{e}_i^{(l)} \cdot \mathbf{E}_l\big|_{z=f^{(l)}(x,y)}, \quad i=1,2, \quad (10a)$$

$$\mathbf{e}_i^{(l)} \cdot \mathbf{H}_{l-1}\big|_{z=f^{(l)}(x,y)} = \mathbf{e}_i^{(l)} \cdot \mathbf{H}_l\big|_{z=f^{(l)}(x,y)}, \quad i=1,2, \quad (10b)$$

where ($\mathbf{e}_1^{(l)}, \mathbf{e}_2^{(l)}, \mathbf{e}_3$) are the ($\mathbf{e}_1, \mathbf{e}_2, \mathbf{e}_3$) defined following Eq. (2), with $\mathbf{e}_1^{(l)}$ and $\mathbf{e}_2^{(l)}$ defined by $z=f^{(l)}(x,y)$, and $\mathbf{e}_3=\mathbf{z}$ shared by all the layer boundaries. Substituting Eq. (5) into Eq. (8) and after careful calculations, the Fourier expansion coefficients of both sides of Eq. (10) can be obtained [57],

$$[\tilde{\mathbf{E}}_{1,l-1}^{(l)}; \tilde{\mathbf{E}}_{2,l-1}^{(l)}; \tilde{\mathbf{H}}_{1,l-1}^{(l)}; \tilde{\mathbf{H}}_{2,l-1}^{(l)}] = [\mathbf{W}'^{(l-1)}_{l-1,+}, \mathbf{W}^{(l)}_{l-1,-}] \begin{bmatrix} \mathbf{u}_{l-1}^{(l-1)} \\ \mathbf{d}_{l-1}^{(l)} \end{bmatrix}, \quad (11a)$$

$$[\tilde{\mathbf{E}}_{1,l}^{(l)}; \tilde{\mathbf{E}}_{2,l}^{(l)}; \tilde{\mathbf{H}}_{1,l}^{(l)}; \tilde{\mathbf{H}}_{2,l}^{(l)}] = [\mathbf{W}_{l,+}^{(l)}, \mathbf{W}'^{(l+1)}_{l,-}] \begin{bmatrix} \mathbf{u}_l^{(l)} \\ \mathbf{d}_l^{(l+1)} \end{bmatrix}, \quad (11b)$$

where the semicolon represents a concatenation of several matrices along their row dimension, and the column vectors $\tilde{\mathbf{E}}_{i,l'}^{(l)}$ and $\tilde{\mathbf{H}}_{i,l'}^{(l)}$ ($i=1, 2$ and $l'=l-1, l$) are defined with their ($m,p$)-th elements being,

$$(\tilde{\mathbf{E}}_{i,l'}^{(l)})_{(m,p),1} = F_{(u',v')}\{\mathbf{e}_i^{(l)} \cdot \mathbf{E}_{l'}\big|_{z=f^{(l)}(x,y)}\}_{(m,p)}, \quad (12a)$$

$$(\tilde{\mathbf{H}}_{i,l'}^{(l)})_{(m,p),1} = F_{(u',v')}\{\mathbf{e}_i^{(l)} \cdot \mathbf{H}_{l'}\big|_{z=f^{(l)}(x,y)}\}_{(m,p)}. \quad (12b)$$

As in Eq. (12), the superscript ~ is used *throughout this paper* to denote a column vector composed of the Fourier expansion coefficients of an electromagnetic field component. In Eq. (11), $\mathbf{u}_l^{(l)}$ and $\mathbf{d}_l^{(l+1)}$ are column vectors with their $r$-th elements being $u_{l,r}^{(l)}$ and $d_{l,r}^{(l+1)}$, respectively, and detailed definitions of the matrices $\mathbf{W}'^{(l-1)}_{l-1,+}$, $\mathbf{W}^{(l)}_{l-1,-}$, $\mathbf{W}_{l,+}^{(l)}$ and $\mathbf{W}'^{(l+1)}_{l,-}$ can be found in Ref. [57]. By using Eq. (11), the equations of boundary condition at $z=f^{(l)}(x,y)$ ($l=1, 2, …, L+1$) in terms of the Fourier expansion coefficients of both sides of Eq. (10) can be finally obtained [57],

$$[\mathbf{W}'^{(l-1)}_{l-1,+}, \mathbf{W}^{(l)}_{l-1,-}] \begin{bmatrix} \mathbf{u}_{l-1}^{(l-1)} \\ \mathbf{d}_{l-1}^{(l)} \end{bmatrix} = [\mathbf{W}_{l,+}^{(l)}, \mathbf{W}'^{(l+1)}_{l,-}] \begin{bmatrix} \mathbf{u}_l^{(l)} \\ \mathbf{d}_l^{(l+1)} \end{bmatrix}, \quad (13)$$

where for $l=1$ or $L+1$ (i.e., at the boundary of the bottom or top semi-infinite layer), it is required to perform a replacement,

$$\mathbf{W}'^{(0)}_{0,+} \mathbf{u}_0^{(0)} \rightarrow \mathbf{W}_{0,+}^{(1)} \mathbf{u}_0^{(1)} \text{ or } \mathbf{W}'^{(L+2)}_{L+1,-} \mathbf{d}_{L+1}^{(L+2)} \rightarrow \mathbf{W}_{L+1,-}^{(L+1)} \mathbf{d}_{L+1}^{(L+1)}, \quad (14)$$

respectively.

To solve the set of linear equations (13) with the modal coefficients $\mathbf{u}_l^{(l)}$ and $\mathbf{d}_l^{(l+1)}$ as the unknowns, a modified scattering-matrix algorithm in Ref. [57] can be used to ensure the numerical stability. In the algorithm, a scattering matrix $\mathbf{S}_l$ relating the modal coefficients in layer 0 and those in layer $l$ ($l=0, 1, 2, …, L+1$) is defined such that,

$$\begin{bmatrix} \mathbf{u}_l^{(l)} \\ \mathbf{d}_0^{(1)} \end{bmatrix} = \mathbf{S}_l \begin{bmatrix} \mathbf{u}_0^{(0)} \\ \mathbf{d}_l^{(l+1)} \end{bmatrix}. \qquad (15)$$

Then started by $\mathbf{S}_0=\mathbf{I}$ being an identity matrix, the algorithm can solve the $\mathbf{S}_l$ iteratively for $l=1, 2, …, L+1$. After solving the $\mathbf{S}_{L+1}$ and setting $\mathbf{u}_0^{(0)}$ and $\mathbf{d}_{L+1}^{(L+2)}$ [with replacement (14)] as the excitation condition, the $\mathbf{d}_0^{(1)}$ and $\mathbf{u}_{L+1}^{(L+1)}$ (and resultantly, all the modal coefficients in all layers) can be obtained.

## III. CFMM UNDER NEBC

In the CFMM under NEBC, the NEBC is set at the boundaries $z=f^{(l)}(x,y)$ ($l=1, 2, …, L+1$) between layers $l-1$ and $l$. The NEBCs satisfied by the tangential electric-field and magnetic-field components are expressed as [11],

$$[\![\mathbf{E}_\parallel]\!] = -\nabla_\parallel(d_\perp [\![E_\perp]\!]), \qquad (16a)$$

$$[\![\mathbf{H}_\parallel]\!] = i\omega d_\parallel [\![\mathbf{D}_\parallel]\!] \times \mathbf{n}, \qquad (16b)$$

where $d_\perp$ and $d_\parallel$ are the Feibelman $d$-parameters, $\mathbf{D}=\boldsymbol{\varepsilon}\cdot\mathbf{E}$, $\nabla_\parallel$ is the surface differential operator [29,58], $[\![\mathbf{F}]\!]=\mathbf{F}^+ - \mathbf{F}^-$ denotes the discontinuity of field $\mathbf{F}$ at $z=f^{(l)}(x,y)$, with $\mathbf{F}^+$ and $\mathbf{F}^-$ denoting the values of field $\mathbf{F}$ at $z=f^{(l)}(x,y)$ on the sides of layers $l$ and $l-1$, respectively, $F_\perp=\mathbf{n}\cdot\mathbf{F}$ and $\mathbf{F}_\parallel=\mathbf{F}-\mathbf{n}F_\perp$ represent the normal and tangential components of $\mathbf{F}$, respectively, and $\mathbf{n}$ is the unit normal vector on $z=f^{(l)}(x,y)$ pointing from layer $l-1$ to layer $l$. Here note that if $\mathbf{n}$ points from dielectric to metal [for instance, on the water-gold interface at $z=f^{(2)}(x,y)$ as shown in Fig. 1(b)], then the commonly used $d$-parameter values (corresponding to $\mathbf{n}$ that points from metal to dielectric) should be multiplied by $-1$.

Equation (16) is written using the international system of units. As mentioned after Eq. (3), in this paper we use the Gaussian system of units for the convenience of numerical implementation, so the $\omega$ in Eq. (16b) actually becomes the wavenumber $k_0$ in the vacuum.

In subsections A and B, respectively starting from NEBC (16a) and (16b) which are in a form independent of coordinate system, we will derive the corresponding NEBC under the curvilinear coordinate system of $(u,v,w)$ defined by Eq. (1), which is expressed in terms of the covariant components of electromagnetic fields solved from Eq. (2). Then in subsections C and D, the NEBC under $(u,v,w)$ is projected upon the Fourier basis, which then yields the equations of NEBC at $z=f^{(l)}(x,y)$ in terms of the Fourier expansion coefficients (i.e., the Fourier representation of NEBC). Finally, in subsection E, the CFMM under NEBC is built up.

### A. Tangential electric-field NEBC under the curvilinear coordinate system of CFMM

Equation (16a) is equivalent to,
$$\mathbf{n}\times[\![\mathbf{E}]\!] = -\mathbf{n}\times[\nabla_\parallel(d_\perp[\![E_\perp]\!])], \qquad (17)$$
where there are,

$$\mathbf{n}\times[\![\mathbf{E}]\!] = \frac{\mathbf{e}^3}{|\mathbf{e}^3|}\times[(\mathbf{e}^1\mathbf{e}_1+\mathbf{e}^2\mathbf{e}_2+\mathbf{e}^3\mathbf{e}_3)\cdot[\![\mathbf{E}]\!]]$$
$$= \frac{\mathbf{e}^3}{|\mathbf{e}^3|}\times(\mathbf{e}^1[\![E_1]\!]+\mathbf{e}^2[\![E_2]\!]+\mathbf{e}^3[\![E_3]\!]) \qquad (18a)$$
$$= \frac{1}{|\mathbf{e}^3|}\left(\frac{1}{V}\mathbf{e}_2[\![E_1]\!] - \frac{1}{V}\mathbf{e}_1[\![E_2]\!]\right),$$

$$-\mathbf{n}\times[\nabla_\parallel(d_\perp[\![E_\perp]\!])]$$
$$= -\frac{\mathbf{e}^3}{|\mathbf{e}^3|}\times[\nabla_\parallel(d_\perp[\![E_\perp]\!])] \qquad (18b)$$
$$= -\frac{1}{|\mathbf{e}^3|}\frac{1}{V}\left[\frac{\partial(d_\perp[\![E_\perp]\!])}{\partial u}\mathbf{e}_2 - \frac{\partial(d_\perp[\![E_\perp]\!])}{\partial v}\mathbf{e}_1\right].$$

Equation (18a) is obtained by using $\mathbf{n}=\mathbf{e}^3/|\mathbf{e}^3|$ and $\mathbf{e}^1\mathbf{e}_1+\mathbf{e}^2\mathbf{e}_2+\mathbf{e}^3\mathbf{e}_3$ being an identity tensor (for the first equality), $E_i=\mathbf{e}_i\cdot\mathbf{E}$ (the second equality), and $\mathbf{e}^i\times\mathbf{e}^j=\mathbf{e}^k/V$ for $(i,j,k)=(1,2,3), (2,3,1), (3,1,2)$ (the third equality). Equation (18b) can be obtained by calculating the tangential components of surface gradient under the curvilinear coordinate system of CFMM (which is not orthogonal in general). Substituting Eq. (18) into Eq. (17), one can obtain,

$$[\![E_1]\!] = -\frac{\partial(d_\perp[\![E_\perp]\!])}{\partial u}, \qquad (19a)$$

$$[\![E_2]\!] = -\frac{\partial(d_\perp[\![E_\perp]\!])}{\partial v}. \qquad (19b)$$

In Eq. (19), there is,

$$[\![E_\perp]\!] = \mathbf{n}\cdot[\![\mathbf{E}]\!]$$
$$= \frac{\mathbf{e}^3}{|\mathbf{e}^3|}\cdot(\mathbf{e}^1\mathbf{e}_1+\mathbf{e}^2\mathbf{e}_2+\mathbf{e}^3\mathbf{e}_3)\cdot[\![\mathbf{E}]\!]$$
$$= \frac{\mathbf{e}^3}{|\mathbf{e}^3|}\cdot(\mathbf{e}^1[\![E_1]\!]+\mathbf{e}^2[\![E_2]\!]+\mathbf{e}^3[\![E_3]\!]) \qquad (20)$$
$$= \frac{1}{|\mathbf{e}^3|}(g^{3,1}[\![E_1]\!]+g^{3,2}[\![E_2]\!]+g^{3,3}[\![E_3]\!]),$$

where $g^{i,j}=\mathbf{e}^i\cdot\mathbf{e}^j$ ($i,j=1, 2, 3$) is the contravariant component of the identity metric tensor. Substituting Eq. (20) into Eq. (19) and applying Eq. (4), we can finally obtain,

$$E_1^- + g_1(u')\frac{1}{k_0}\frac{\partial E_{123}^-}{\partial u'} = E_1^+ + g_1(u')\frac{1}{k_0}\frac{\partial E_{123}^+}{\partial u'}, \qquad (21a)$$

$$E_2^- + g_2(v')\frac{1}{k_0}\frac{\partial E_{123}^-}{\partial v'} = E_2^+ + g_2(v')\frac{1}{k_0}\frac{\partial E_{123}^+}{\partial v'}, \qquad (21b)$$

where we define,

$$E_{123}^{\pm} = d'_{\perp} \frac{g^{3,1}}{|\mathbf{e}^3|} E_1^{\pm} + d'_{\perp} \frac{g^{3,2}}{|\mathbf{e}^3|} E_2^{\pm} + d'_{\perp} \frac{g^{3,3}}{|\mathbf{e}^3|} E_3^{\pm}, \quad (22)$$

with $d'_{\perp} = k_0 d_{\perp}$ defined as a dimensionless $d$-parameter, and $g^{3,3}/|\mathbf{e}^3| = |\mathbf{e}^3|$. Equation (21) along with Eq. (22) is the main result of this subsection, i.e., the tangential electric-field NEBC under the curvilinear coordinate system $(u^1, u^2, u^3) = (u, v, w)$ of CFMM.

## B. Tangential magnetic-field NEBC under the curvilinear coordinate system of CFMM

Equation (16b) is equivalent to,
$$\mathbf{e}_i \cdot [\![\mathbf{H}]\!] = i d'_{\parallel} \mathbf{e}_i \cdot ([\![\mathbf{D}]\!] \times \mathbf{n}), \ i = 1, 2, \quad (23)$$

where $d'_{\parallel} = k_0 d_{\parallel}$ is defined as a dimensionless $d$-parameter, and $\omega$ is replaced by $k_0$ for the Gaussian system of units used in this paper. In Eq. (23), there is,

$$\begin{aligned}
& i d'_{\parallel} \mathbf{e}_1 \cdot ([\![\mathbf{D}]\!] \times \mathbf{n}) \\
& \stackrel{1}{=} -i d'_{\parallel} \mathbf{e}_1 \cdot \left\{ \frac{\mathbf{e}^3}{|\mathbf{e}^3|} \times [(\mathbf{e}_1 \mathbf{e}^1 + \mathbf{e}_2 \mathbf{e}^2 + \mathbf{e}_3 \mathbf{e}^3) \cdot [\![\mathbf{D}]\!]] \right\} \\
& \stackrel{2}{=} -i d'_{\parallel} \mathbf{e}_1 \cdot \left[ \frac{\mathbf{e}^3}{|\mathbf{e}^3|} \times (\mathbf{e}_1 [\![D^1]\!] + \mathbf{e}_2 [\![D^2]\!] + \mathbf{e}_3 [\![D^3]\!]) \right] \\
& \stackrel{3}{=} i d'_{\parallel} \frac{\mathbf{e}^3}{|\mathbf{e}^3|} \cdot (\mathbf{e}_1 \times \mathbf{e}_2) [\![D^2]\!] - i d'_{\parallel} \frac{\mathbf{e}^3}{|\mathbf{e}^3|} \cdot (\mathbf{e}_3 \times \mathbf{e}_1) [\![D^3]\!] \\
& \stackrel{4}{=} i d'_{\parallel} \frac{\mathbf{e}^3}{|\mathbf{e}^3|} \cdot (V\mathbf{e}^3) [\![D^2]\!] - i d'_{\parallel} \frac{\mathbf{e}^3}{|\mathbf{e}^3|} \cdot (V\mathbf{e}^2) [\![D^3]\!] \\
& \stackrel{5}{=} i d'_{\parallel} \frac{g^{3,3}}{|\mathbf{e}^3|} [\![D_2]\!] - i d'_{\parallel} \frac{g^{3,2}}{|\mathbf{e}^3|} [\![D_3]\!],
\end{aligned} \quad (24)$$

which is obtained by using $\mathbf{n} = \mathbf{e}^3/|\mathbf{e}^3|$ and $\mathbf{e}_1 \mathbf{e}^1 + \mathbf{e}_2 \mathbf{e}^2 + \mathbf{e}_3 \mathbf{e}^3$ being an identity tensor (for equality 1), $D^i = \mathbf{e}^i \cdot \mathbf{D}$ (equality 2), $\mathbf{e}_i \times \mathbf{e}_j = V\mathbf{e}^k$ for $(i,j,k) = (1,2,3), (2,3,1), (3,1,2)$ (equality 4), $g^{i,j} = \mathbf{e}^i \cdot \mathbf{e}^j$ and $D_i = VD^i$ (equality 5). Note that $D_i \neq \mathbf{e}_i \cdot \mathbf{D}$ for this definition of $D_i$. According to Eq. (2b), the $D_i$ is given by the constitutive equation,

$$\begin{bmatrix} D_1 \\ D_2 \\ D_3 \end{bmatrix} = \begin{bmatrix} \varepsilon_{1,1} & \varepsilon_{1,2} & \varepsilon_{1,3} \\ \varepsilon_{2,1} & \varepsilon_{2,2} & \varepsilon_{2,3} \\ \varepsilon_{3,1} & \varepsilon_{3,2} & \varepsilon_{3,3} \end{bmatrix} \begin{bmatrix} E_1 \\ E_2 \\ E_3 \end{bmatrix}, \quad (25)$$

where we define $\varepsilon_{i,j} = V \varepsilon^{i,j}$. Also note that $\varepsilon_{i,j} \neq \mathbf{e}_i \cdot \boldsymbol{\varepsilon} \cdot \mathbf{e}_j$ for this definition of $\varepsilon_{i,j}$. Similar to Eq. (24), there is,

$$i d'_{\parallel} \mathbf{e}_2 \cdot ([\![\mathbf{D}]\!] \times \mathbf{n}) = -i d'_{\parallel} \frac{g^{3,3}}{|\mathbf{e}^3|} [\![D_1]\!] + i d'_{\parallel} \frac{g^{3,1}}{|\mathbf{e}^3|} [\![D_3]\!]. \quad (26)$$

Substituting Eqs. (24) and (26) into Eq. (23), one can obtain,
$$H_1^- + D_{23}^- = H_1^+ + D_{23}^+, \quad (27a)$$
$$H_2^- + D_{13}^- = H_2^+ + D_{13}^+, \quad (27b)$$

where $H_i = \mathbf{e}_i \cdot \mathbf{H}$ and we define,
$$D_{23}^{\pm} = -i d'_{\parallel} \frac{g^{3,3}}{|\mathbf{e}^3|} D_2^{\pm} + i d'_{\parallel} \frac{g^{3,2}}{|\mathbf{e}^3|} D_3^{\pm}, \quad (28a)$$

$$D_{13}^{\pm} = i d'_{\parallel} \frac{g^{3,3}}{|\mathbf{e}^3|} D_1^{\pm} - i d'_{\parallel} \frac{g^{3,1}}{|\mathbf{e}^3|} D_3^{\pm}. \quad (28b)$$

Equation (27) along with Eq. (28) is the main result of this subsection, i.e., the tangential magnetic-field NEBC under the curvilinear coordinate system $(u^1, u^2, u^3) = (u, v, w)$ of CFMM.

## C. Fourier representation of the tangential electric-field NEBC under the curvilinear coordinate system of CFMM

By applying the operator $F_{(u',v')}\{\cdot\}_{(m,p)}$ of calculating the Fourier expansion coefficient to both sides of Eq. (21), one can obtain,
$$\tilde{\mathbf{E}}_1^- + i\boldsymbol{\alpha}\tilde{\mathbf{E}}_{123}^- = \tilde{\mathbf{E}}_1^+ + i\boldsymbol{\alpha}\tilde{\mathbf{E}}_{123}^+, \quad (29a)$$
$$\tilde{\mathbf{E}}_2^- + i\boldsymbol{\beta}\tilde{\mathbf{E}}_{123}^- = \tilde{\mathbf{E}}_2^+ + i\boldsymbol{\beta}\tilde{\mathbf{E}}_{123}^+, \quad (29b)$$

where we define column vectors $\tilde{\mathbf{E}}_i^{\pm}$ ($i=1, 2, 3$) and $\tilde{\mathbf{E}}_{123}^{\pm}$ with elements,
$$(\tilde{\mathbf{E}}_i^{\pm})_{(m,p),1} = F_{(u',v')}\{E_i^{\pm}\}_{(m,p)}, \quad (30a)$$
$$(\tilde{\mathbf{E}}_{123}^{\pm})_{(m,p),1} = F_{(u',v')}\{E_{123}^{\pm}\}_{(m,p)}, \quad (30b)$$

and detailed definitions of matrices $\boldsymbol{\alpha}$ and $\boldsymbol{\beta}$ can be found in Supplemental Material Sec. S1 [60] [Eq. (S1.34)]. The setting of $g_1(u')$ and $g_2(v')$ to be continuous functions has been used to apply the Laurent's rule [35] for deriving Eq. (29).

The next key step is to obtain the $\tilde{\mathbf{E}}_{123}^{\pm}$ expressed in terms of the Fourier expansion coefficients of the tangential covariant field components $(E_1^{\pm}, E_2^{\pm}, H_1^{\pm}, H_2^{\pm})$, i.e., the Fourier representation of $\tilde{\mathbf{E}}_{123}^{\pm}$. However, the problem is that in Eq. (22), the $d'_{\perp}$ and $g^{3,j}/|\mathbf{e}^3|$ ($j=1, 2, 3$) are in general allowed to be discontinuous with respect to $u'$ and $v'$, and the field components $E_1^{\pm}$ and $E_2^{\pm}$ are also possibly discontinuous with respect to $u'$ or $v'$, so that the *correct Fourier factorization rules* [35] cannot be applied to Eq. (22) directly. To address the problem, by employing the constitutive Eq. (25) or its Fourier representation, the discontinuous field components (or their Fourier expansion coefficients) in Eq. (22) can be expressed in terms of the field components (or their Fourier expansion coefficients) that are continuous with respect to either $u'$ or $v'$, so that the Laurent's rule [35] is applicable, and we finally obtain (see the detailed derivation in Supplemental Material Sec. S2 [60]),

$$\tilde{\mathbf{E}}_{123}^{\pm} = \mathbf{Q}_{5,1}^{\pm}\tilde{\mathbf{E}}_1^{\pm} + \mathbf{Q}_{5,2}^{\pm}\tilde{\mathbf{E}}_2^{\pm} + \mathbf{Q}_{5,3}^{\pm}\tilde{\mathbf{H}}_1^{\pm} + \mathbf{Q}_{5,4}^{\pm}\tilde{\mathbf{H}}_2^{\pm}, \quad (31)$$

where the detailed definitions of the matrices $\mathbf{Q}_{5,i}^{\pm}$ ($i=1, 2, 3, 4$) can be found in Eq. (S2.10), and we define column vectors $\tilde{\mathbf{H}}_i^{\pm}$ ($i=1, 2, 3$) with elements,
$$(\tilde{\mathbf{H}}_i^{\pm})_{(m,p),1} = F_{(u',v')}\{H_i^{\pm}\}_{(m,p)}. \quad (32)$$

To derive Eq. (31) [i.e., Eq. (S2.9)], the Fourier representation of the Maxwell' equation (2b) is also used to eliminate the Fourier expansion coefficients $\tilde{\mathbf{E}}_3^\pm$ of $E_3^\pm$ in Eq. (22), which then results in the presence of the $\tilde{\mathbf{H}}_i^\pm$ in Eq. (31).

Substituting Eq. (31) into Eq. (29), one can obtain,

$$\mathbf{Q}_{6,1}^-\tilde{\mathbf{E}}_1^- + \mathbf{Q}_{6,2}^-\tilde{\mathbf{E}}_2^- + \mathbf{Q}_{6,3}^-\tilde{\mathbf{H}}_1^- + \mathbf{Q}_{6,4}^-\tilde{\mathbf{H}}_2^-$$
$$= \mathbf{Q}_{6,1}^+\tilde{\mathbf{E}}_1^+ + \mathbf{Q}_{6,2}^+\tilde{\mathbf{E}}_2^+ + \mathbf{Q}_{6,3}^+\tilde{\mathbf{H}}_1^+ + \mathbf{Q}_{6,4}^+\tilde{\mathbf{H}}_2^+, \quad (33a)$$

$$\mathbf{Q}_{7,1}^-\tilde{\mathbf{E}}_1^- + \mathbf{Q}_{7,2}^-\tilde{\mathbf{E}}_2^- + \mathbf{Q}_{7,3}^-\tilde{\mathbf{H}}_1^- + \mathbf{Q}_{7,4}^-\tilde{\mathbf{H}}_2^-$$
$$= \mathbf{Q}_{7,1}^+\tilde{\mathbf{E}}_1^+ + \mathbf{Q}_{7,2}^+\tilde{\mathbf{E}}_2^+ + \mathbf{Q}_{7,3}^+\tilde{\mathbf{H}}_1^+ + \mathbf{Q}_{7,4}^+\tilde{\mathbf{H}}_2^+, \quad (33b)$$

where the matrices $\mathbf{Q}_{6,i}^\pm$ and $\mathbf{Q}_{7,i}^\pm$ ($i$=1, 2, 3, 4) are given by,

$$\mathbf{Q}_{6,1}^\pm = \mathbf{I} + i\alpha\mathbf{Q}_{5,1}^\pm, \quad \mathbf{Q}_{6,2}^\pm = i\alpha\mathbf{Q}_{5,2}^\pm,$$
$$\mathbf{Q}_{6,3}^\pm = i\alpha\mathbf{Q}_{5,3}^\pm, \quad \mathbf{Q}_{6,4}^\pm = i\alpha\mathbf{Q}_{5,4}^\pm, \quad (34a)$$

$$\mathbf{Q}_{7,1}^\pm = i\beta\mathbf{Q}_{5,1}^\pm, \quad \mathbf{Q}_{7,2}^\pm = \mathbf{I} + i\beta\mathbf{Q}_{5,2}^\pm,$$
$$\mathbf{Q}_{7,3}^\pm = i\beta\mathbf{Q}_{5,3}^\pm, \quad \mathbf{Q}_{7,4}^\pm = i\beta\mathbf{Q}_{5,4}^\pm. \quad (34b)$$

Equation (33) is the main result of this subsection, i.e., the Fourier representation of the tangential electric-field NEBC [Eq. (21)] under the curvilinear coordinate system of CFMM.

### D. Fourier representation of the tangential magnetic-field NEBC under the curvilinear coordinate system of CFMM

By applying the operator $F_{(u',v')}\{\}_{(m,p)}$ of calculating the Fourier expansion coefficient to both sides of Eq. (27), one can obtain,

$$\tilde{\mathbf{H}}_1^- + \tilde{\mathbf{D}}_{23}^- = \tilde{\mathbf{H}}_1^+ + \tilde{\mathbf{D}}_{23}^+, \quad (35a)$$
$$\tilde{\mathbf{H}}_2^- + \tilde{\mathbf{D}}_{13}^- = \tilde{\mathbf{H}}_2^+ + \tilde{\mathbf{D}}_{13}^+, \quad (35b)$$

where the column vectors $\tilde{\mathbf{D}}_{23}^\pm$ and $\tilde{\mathbf{D}}_{13}^\pm$ are defined with elements,

$$(\tilde{\mathbf{D}}_{23}^\pm)_{(m,p),1} = F_{(u',v')}\{D_{23}^\pm\}_{(m,p)}, \quad (36a)$$
$$(\tilde{\mathbf{D}}_{13}^\pm)_{(m,p),1} = F_{(u',v')}\{D_{13}^\pm\}_{(m,p)}. \quad (36b)$$

The next key step is to obtain the $\tilde{\mathbf{D}}_{23}^\pm$ and $\tilde{\mathbf{D}}_{13}^\pm$ expressed in terms of the Fourier expansion coefficients of the tangential covariant field components $(E_1^\pm, E_2^\pm, H_1^\pm, H_2^\pm)$, i.e., the Fourier representation of $\tilde{\mathbf{D}}_{23}^\pm$ and $\tilde{\mathbf{D}}_{13}^\pm$. However, the problem is that in Eq. (28), the $d_\parallel'$ and $g^{3,j}/|\mathbf{e}^3|$ ($j$=1, 2, 3) are in general allowed to be discontinuous with respect to $u'$ and $v'$, and the field components $D_i^\pm$ ($i$=1, 2, 3) are also possibly discontinuous with respect to $u'$ or $v'$, so that the *correct Fourier factorization rules* [35] cannot be applied to Eq. (28) directly. This problem can be treated similar to Eq. (31), and we finally obtain (see the detailed derivation in Supplemental Material Secs. S3 and S4 [60]),

$$\tilde{\mathbf{D}}_{23}^\pm = \mathbf{Q}_{12,1}^\pm\tilde{\mathbf{E}}_1^\pm + \mathbf{Q}_{12,2}^\pm\tilde{\mathbf{E}}_2^\pm + \mathbf{Q}_{12,3}^\pm\tilde{\mathbf{H}}_1^\pm + \mathbf{Q}_{12,4}^\pm\tilde{\mathbf{H}}_2^\pm, \quad (37a)$$
$$\tilde{\mathbf{D}}_{13}^\pm = \mathbf{Q}_{18,1}^\pm\tilde{\mathbf{E}}_1^\pm + \mathbf{Q}_{18,2}^\pm\tilde{\mathbf{E}}_2^\pm + \mathbf{Q}_{18,3}^\pm\tilde{\mathbf{H}}_1^\pm + \mathbf{Q}_{18,4}^\pm\tilde{\mathbf{H}}_2^\pm, \quad (37b)$$

where the detailed definitions of the matrices $\mathbf{Q}_{12,i}^\pm$ and $\mathbf{Q}_{18,i}^\pm$ ($i$=1, 2, 3, 4) can be found in Eqs. (S3.5) and (S4.5), respectively. Substituting Eq. (37) into Eq. (35), one can obtain,

$$\mathbf{Q}_{13,1}^-\tilde{\mathbf{E}}_1^- + \mathbf{Q}_{13,2}^-\tilde{\mathbf{E}}_2^- + \mathbf{Q}_{13,3}^-\tilde{\mathbf{H}}_1^- + \mathbf{Q}_{13,4}^-\tilde{\mathbf{H}}_2^-$$
$$= \mathbf{Q}_{13,1}^+\tilde{\mathbf{E}}_1^+ + \mathbf{Q}_{13,2}^+\tilde{\mathbf{E}}_2^+ + \mathbf{Q}_{13,3}^+\tilde{\mathbf{H}}_1^+ + \mathbf{Q}_{13,4}^+\tilde{\mathbf{H}}_2^+, \quad (38a)$$

$$\mathbf{Q}_{19,1}^-\tilde{\mathbf{E}}_1^- + \mathbf{Q}_{19,2}^-\tilde{\mathbf{E}}_2^- + \mathbf{Q}_{19,3}^-\tilde{\mathbf{H}}_1^- + \mathbf{Q}_{19,4}^-\tilde{\mathbf{H}}_2^-$$
$$= \mathbf{Q}_{19,1}^+\tilde{\mathbf{E}}_1^+ + \mathbf{Q}_{19,2}^+\tilde{\mathbf{E}}_2^+ + \mathbf{Q}_{19,3}^+\tilde{\mathbf{H}}_1^+ + \mathbf{Q}_{19,4}^+\tilde{\mathbf{H}}_2^+, \quad (38b)$$

where the matrices $\mathbf{Q}_{13,i}^\pm$ and $\mathbf{Q}_{19,i}^\pm$ ($i$=1, 2, 3, 4) are given by,

$$\mathbf{Q}_{13,1}^\pm = \mathbf{Q}_{12,1}^\pm, \quad \mathbf{Q}_{13,2}^\pm = \mathbf{Q}_{12,2}^\pm,$$
$$\mathbf{Q}_{13,3}^\pm = \mathbf{I} + \mathbf{Q}_{12,3}^\pm, \quad \mathbf{Q}_{13,4}^\pm = \mathbf{Q}_{12,4}^\pm, \quad (39a)$$

$$\mathbf{Q}_{19,1}^\pm = \mathbf{Q}_{18,1}^\pm, \quad \mathbf{Q}_{19,2}^\pm = \mathbf{Q}_{18,2}^\pm,$$
$$\mathbf{Q}_{19,3}^\pm = \mathbf{Q}_{18,3}^\pm, \quad \mathbf{Q}_{19,4}^\pm = \mathbf{I} + \mathbf{Q}_{18,4}^\pm. \quad (39b)$$

Equation (38) is the main result of this subsection, i.e., the Fourier representation of the tangential magnetic-field NEBC [Eq. (27)] under the curvilinear coordinate system of CFMM.

### E. CFMM under NEBC

Equations (33) and (38) can be rewritten as,

$$\mathbf{Q}_{20}^-\tilde{\boldsymbol{\psi}}_{12}^- = \mathbf{Q}_{20}^+\tilde{\boldsymbol{\psi}}_{12}^+, \quad (40)$$

where the column vectors $\tilde{\boldsymbol{\psi}}_{12}^\pm$ and matrices $\mathbf{Q}_{20}^\pm$ are defined as,

$$\tilde{\boldsymbol{\psi}}_{12}^\pm = [\tilde{\mathbf{E}}_1^\pm; \tilde{\mathbf{E}}_2^\pm; \tilde{\mathbf{H}}_1^\pm; \tilde{\mathbf{H}}_2^\pm], \quad (41a)$$

$$\mathbf{Q}_{20}^\pm = \begin{bmatrix} \mathbf{Q}_{6,1}^\pm & \mathbf{Q}_{6,2}^\pm & \mathbf{Q}_{6,3}^\pm & \mathbf{Q}_{6,4}^\pm \\ \mathbf{Q}_{7,1}^\pm & \mathbf{Q}_{7,2}^\pm & \mathbf{Q}_{7,3}^\pm & \mathbf{Q}_{7,4}^\pm \\ \mathbf{Q}_{13,1}^\pm & \mathbf{Q}_{13,2}^\pm & \mathbf{Q}_{13,3}^\pm & \mathbf{Q}_{13,4}^\pm \\ \mathbf{Q}_{19,1}^\pm & \mathbf{Q}_{19,2}^\pm & \mathbf{Q}_{19,3}^\pm & \mathbf{Q}_{19,4}^\pm \end{bmatrix}. \quad (41b)$$

For the notations in Eqs. (40) and (11), there are,

$$\tilde{\boldsymbol{\psi}}_{12}^- = [\tilde{\mathbf{E}}_{1,l-1}^{(l)}; \tilde{\mathbf{E}}_{2,l-1}^{(l)}; \tilde{\mathbf{H}}_{1,l-1}^{(l)}; \tilde{\mathbf{H}}_{2,l-1}^{(l)}]. \quad (42a)$$

$$\tilde{\boldsymbol{\psi}}_{12}^+ = [\tilde{\mathbf{E}}_{1,l}^{(l)}; \tilde{\mathbf{E}}_{2,l}^{(l)}; \tilde{\mathbf{H}}_{1,l}^{(l)}; \tilde{\mathbf{H}}_{2,l}^{(l)}], \quad (42b)$$

Then by substituting Eq. (11) into Eq. (40), one can obtain,

$$\mathbf{Q}_{20}^-[\mathbf{W}_{l-1,+}'^{(l-1)}, \mathbf{W}_{l-1,-}^{(l)}]\begin{bmatrix}\mathbf{u}_{l-1}^{(l-1)}\\\mathbf{d}_{l-1}^{(l)}\end{bmatrix} = \mathbf{Q}_{20}^+[\mathbf{W}_{l,+}^{(l)}, \mathbf{W}_{l,-}'^{(l+1)}]\begin{bmatrix}\mathbf{u}_l^{(l)}\\\mathbf{d}_l^{(l+1)}\end{bmatrix}, \quad (43)$$

which are the equations of NEBC at $z=f^{(l)}(x,y)$ ($l$=1, 2, …, $L$+1) in terms of the Fourier expansion coefficients (i.e., the Fourier representation of NEBC). By solving Eq. (43), the unknown modal coefficients $\mathbf{u}_l^{(l)}$ and $\mathbf{d}_l^{(l+1)}$ ($l$=0, 1, 2, …, $L$+1) can be obtained. Remarkably, a comparison between Eq. (43) of NEBC and the corresponding Eq. (13) of CEBC shows that the former can become the latter simply after the following replacement,

$$\mathbf{Q}_{20}^{-}[\mathbf{W}_{l-1,+}'^{(l-1)}, \mathbf{W}_{l-1,-}^{(l)}] \rightarrow [\mathbf{W}_{l-1,+}'^{(l-1)}, \mathbf{W}_{l-1,-}'^{(l)}], \quad (44a)$$

$$\mathbf{Q}_{20}^{+}[\mathbf{W}_{l,+}^{(l)}, \mathbf{W}_{l,-}^{(l+1)}] \rightarrow [\mathbf{W}_{l,+}'^{(l)}, \mathbf{W}_{l,-}'^{(l+1)}], \quad (44b)$$

which implies that after the replacement (44), the modified scattering-matrix algorithm in Ref. [57] [see Eq. (15) and the relevant statements] can be applied directly for solving Eq. (43), so that the modal coefficients $\mathbf{u}_l^{(l)}$ and $\mathbf{d}_l^{(l+1)}$ [and resultantly, the electromagnetic field as expressed in Eq. (5)] can be obtained.

## IV. NUMERICAL TEST

To test the validity and accuracy of the proposed CFMM under NEBC, which is implemented with an in-house software [61], we consider the numerical example of a single gold nanosphere (radius $R$=20nm) on a flat gold substrate (with a nanogap of size $g$) in water [as sketched in Fig. 1(a)] excited by a normally-incident $x$-polarized plane wave of wavelength $\lambda$=633nm. This is a typical nanoparticle-on-mirror plasmonic structure. Such structures have advantages of an easy formation of metallic nanogaps down to nanometer sizes, where a significant enhancement of electric field can be achieved [3,4,10,14] along with pronounced nonclassical quantum effects [4,10,11,14].

For applying the CFMM under NEBC, Fig. 1(b) shows that the whole structure can be decomposed into 4 $z$-invariant layers ($l$=0, 1, 2, 3) separated by boundary surfaces $z=f^{(l)}(x,y)$ ($l$=1, 2, 3), where the coordinate origin $O$ is set at the center of the gold nanosphere. Note that if the top (within $z>0$) and bottom ($z<0$) surfaces of the nanosphere are treated as the top and bottom boundaries of the central layer 2, the derivatives $f_x$ and $f_y$ of the profile function $z=f(x,y)$ will take infinitely large values at $z=0$ [i.e. at $P_7$ and $P_8$ in Fig. 1(b)], which will results in numerical instability in the calculation of Eq. (3). To solve this problem, we adopt the treatment in Ref. [57] by instead considering two approximate boundaries of layer 2 along with a medium perturbation of the nanosphere. For the approximate bottom boundary, we consider a conical surface which is tangential to the bottom sphere surface at $P_3$ and $P_4$ (with a small angle $\theta$), and which intersects the sphere's circumscribed cylindrical surface (containing $P_7$ and $P_8$) at $P_2$ and $P_5$, as illustrated in Fig. 1(b). Then the approximate bottom boundary of layer 2 is composed of a planar surface ($P_1P_2$ and $P_5P_6$), the conical surface ($P_2P_3$ and $P_4P_5$) and the sphere surface ($P_3P_4$). Correspondingly, a medium perturbation of the bottom semi-sphere (within $z<0$) is introduced such that the medium in the tiny region surrounded by $P_2P_3P_7$ and $P_4P_5P_8$ is changed from the surrounding medium of water to the nanosphere medium of gold. The approximate top boundary of layer 2 and the corresponding medium perturbation of the top semi-sphere (within $z>0$) are set in a similar way, as illustrated in Fig. 1(b) with points $P_1'$ - $P_6'$, $P_7$ and $P_8$. Thus, for a small value of $\theta$ ($\theta$=5° for the numerical example), the actual nanosphere can be well approximated by the perturbed nanosphere with the approximate boundaries of layer 2, for which the derivatives $f_x$ and $f_y$ of the profile function $z=f(x,y)$ take a finite value everywhere.

For the matched coordinate transformation [45,46], the PMLs [44] and the ASR [42,43], we adopt the setting in Ref. [57] where a numerical example of a single gold nanosphere on a flat gold substrate in air is considered. As shown in Fig. 1(c), the matched coordinate transformation $x=x(u,v)$ and $y=y(u,v)$ is defined such that the inner boundaries $x=\pm w_x/2$ and $y=\pm w_y/2$ of the PMLs and the two circular boundaries of the above defined conical surface are at the curves of $u$ or $v$ being constants (shown by circles), and that there is $(x,y)=(u,v)$ within the region of PMLs. The curves of $u$=constant or $v$=constant are shown by vertically-oriented red curves or horizontally-oriented blue curves, respectively, and are determined with a linear interpolation between the curves of circles [45,46,57]. The ASR for enhancing the spatial resolution and resultantly the convergence [43,57] is applied near the outer circular boundary of the conical surface, where the permittivity is discontinuous and the $f_x$ and $f_y$ take large values (for a small $\theta$) and are also discontinuous.

The NEBC is set at the boundary surfaces $z=f^{(l)}(x,y)$ ($l$=1, 2, 3) as required by the CFMM under NEBC, and the Feibelman $d$-parameters take values $d_\perp=-0.4+0.2i$ nm and $d_\parallel=0.4+0.2i$ nm at the gold-water interfaces (between the gold nanosphere or gold substrate and water) and $d_\perp=d_\parallel=0$ at the virtual water-water interfaces [at $z=f^{(l)}(x,y)$ with $l$=2, 3]. Here the value of $d_\perp$ refers to Refs. [11,29], and commonly there is $|d_\parallel|<<|d_\perp|$ [11,29]. We artificially increase the value of $d_\parallel$ such that $|d_\parallel|=|d_\perp|$, which is to ensure that the impacts of the $d_\parallel$ and $d_\perp$ on the optical response are comparable so as to provide a strong numerical test of the accuracy of the CFMM under NEBC that incorporates both $d$-parameters. Besides, the signs of the real and imaginary parts of $d_\parallel$ are set to be consistent with the $s$-$d$ polarization model [20,29,62]. Note that no NEBC is set at the tiny $z$-invariant gold-water interface [as shown by the $P_2P_2'$ and $P_5P_5'$ in Fig. 1(b)], which causes negligible error as confirmed by the numerical results shown hereafter. The refractive index of gold is 0.1807+2.9970$i$ at the excitation wavelength $\lambda$=633nm [63], and the refractive index of water is 1.33.

Figures 2(a) and 2(b) show the distributions of the dominant electric-field component $|E_z|$ on the gold substrate (i.e., at $z=-R-g$, with the nanogap size $g$=2nm) in water region, which are calculated with the CFMM and FEM for ($d_\perp$=0, $d_\parallel$=0), ($d_\perp$=0, $d_\parallel\neq$0), ($d_\perp\neq$0, $d_\parallel$=0) and ($d_\perp\neq$0, $d_\parallel\neq$0) at the gold-water interfaces, respectively. Figures 2(c) and 2(d) show the distributions of electric-field components $|E_x|$ and $|E_z|$ and magnetic-field component $|H_y|$ in the cross section $y$=0, which are calculated with the CFMM and FEM for $d_\perp\neq$0 and $d_\parallel\neq$0. In Figs. 2(a)-(d), the calculated electric- and magnetic-field components are normalized by the electric- and magnetic-field magnitudes $|\mathbf{E}_{inc}|$ and $|\mathbf{H}_{inc}|$ of the incident plane wave, respectively, and the truncated harmonic orders in the CFMM are set to be $M_x=M_y$=50. Here we use the COMSOL Multiphysics software to perform the FEM, where the NEBC is implemented by introducing a scalar auxiliary potential to ensure the numerical stability [11]. Figures 2(a)-(d) show good agreement between the results of the CFMM and FEM, which confirms the validity and accuracy of the proposed CFMM under NEBC. This agreement also indicates that the error caused by the boundary and medium perturbations in the CFMM [as described in Fig. 1(b)] is negligible, since such perturbations are not introduced in the FEM calculation.

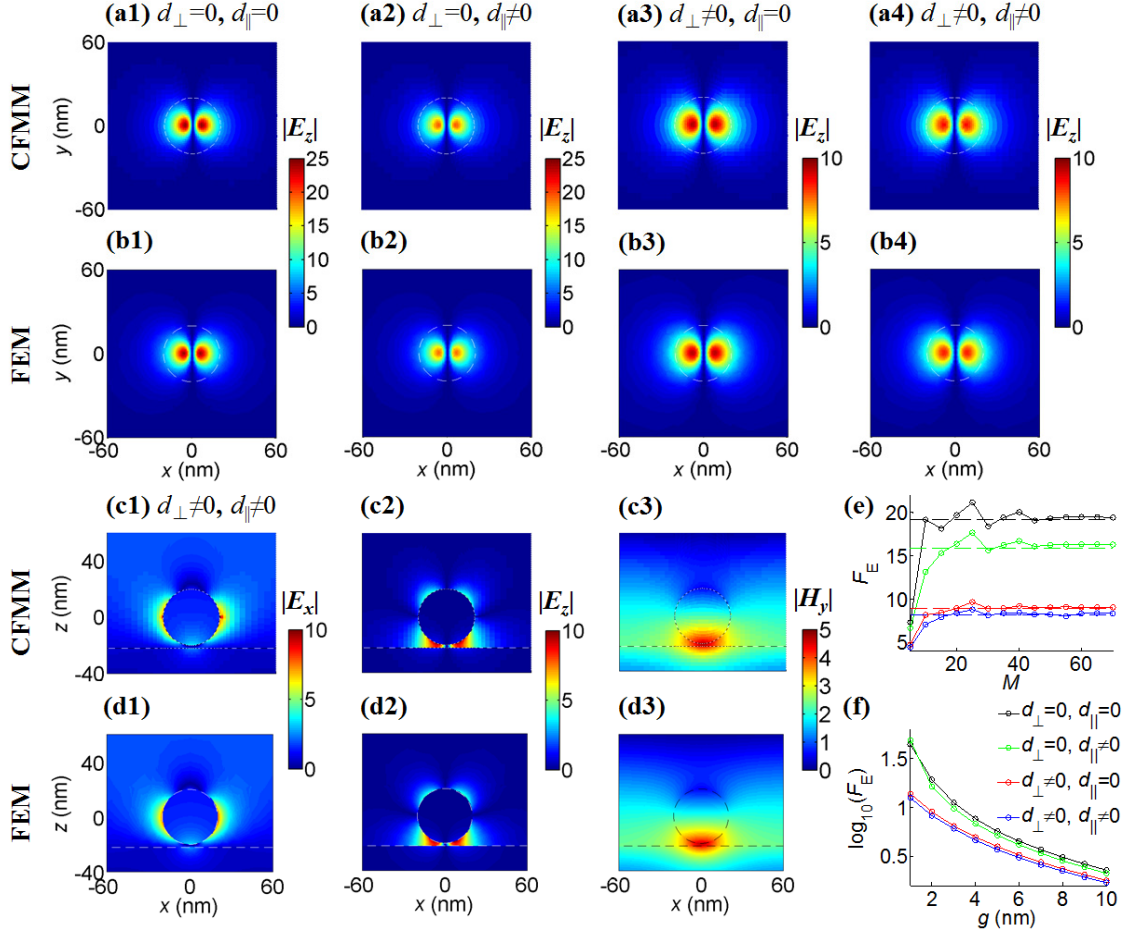

FIG. 2. Results for the numerical example of a single gold nanosphere (radius $R$=20nm) on a flat gold substrate in water excited by a normally-incident $x$-polarized plane wave (with a wavelength $\lambda$=633nm and the electric- and magnetic-field vectors denoted by $\mathbf{E}_{inc}$ and $\mathbf{H}_{inc}$, respectively). **(a), (b)** Distributions of the dominant electric-field component $|E_z|$ (normalized by $|\mathbf{E}_{inc}|$) on the gold substrate (i.e., at $z=-R-g$) in water region. The results are obtained with the CFMM (a1)-(a4) and the FEM (b1)-(b4) for $d_\perp$=0, $d_\parallel$=0 (a1), (b1), $d_\perp$=0, $d_\parallel\neq$0 (a2), (b2), $d_\perp\neq$0, $d_\parallel$=0 (a3), (b3) and $d_\perp\neq$0, $d_\parallel\neq$0 (a4), (b4). **(c), (d)** Distributions of electric-field components $|E_x|$ (c1), (d1) and $|E_z|$ (c2), (d2) (normalized by $|\mathbf{E}_{inc}|$) and magnetic-field component $|H_y|$ (c3), (d3) (normalized by $|\mathbf{H}_{inc}|$) in the cross section $y$=0. The results are obtained with the CFMM (c1)-(c3) and the FEM (d1)-(d3) for $d_\perp\neq$0 and $d_\parallel\neq$0. **(e), (f)** Electric-field enhancement factor $F_E=|E_z|/|\mathbf{E}_{inc}|$ defined at $(x,y,z)=(10\text{nm},0,-R-g)$ in the nanogap (on the gold substrate in water region) plotted as a function of the truncated harmonic order $M_x=M_y=M$ in the CFMM (e) or of the nanogap size $g$ (f). The results are obtained with the CFMM (curves) and the FEM [horizontal dashed lines in (e)] for $d_\perp$=0, $d_\parallel$=0 (black curves), $d_\perp$=0, $d_\parallel\neq$0 (green), $d_\perp\neq$0, $d_\parallel$=0 (red) and $d_\perp\neq$0, $d_\parallel\neq$0 (blue). For (a)-(d), we set $M_x=M_y$=50 in the CFMM and $g$=2nm, and the superimposed white or black dashed lines show the boundaries of the gold nanosphere or substrate. For (e) and (f), we set $g$=2nm and $M_x=M_y$=50, respectively. We set $\theta$=5° for all the results of CFMM.

Figures 2(a)-(d) show that there exists a significant enhancement of electric field in the nanogap between the gold nanosphere and substrate [4,10,14], which is apparently affected by either of the $d$-parameters $d_\perp$ and $d_\parallel$ in view of the difference between the results of $d_\perp \neq 0$ and $d_\perp=0$ (for fixed $d_\parallel \neq 0$ or $d_\parallel=0$) or between the results of $d_\parallel \neq 0$ and $d_\parallel=0$ (for fixed $d_\perp \neq 0$ or $d_\perp=0$). Figures 2(e) and 2(f) provide an electric-field enhancement factor $F_E=|E_z|/|\mathbf{E}_{inc}|$ defined at $(x,y,z)=(10\text{nm},0,-R-g)$ in the nanogap (on the gold substrate in water region) plotted as a function of the truncated harmonic order $M_x=M_y=M$ in the CFMM (with $g$=2nm) or of the nanogap size $g$ (with $M_x=M_y=50$), respectively. The results are obtained with the CFMM (curves) and the FEM [horizontal dashed lines in Fig. 2(e)] for $d_\perp=0$, $d_\parallel=0$ (black curves), $d_\perp=0$, $d_\parallel \neq 0$ (green), $d_\perp \neq 0$, $d_\parallel=0$ (red) and $d_\perp \neq 0$, $d_\parallel \neq 0$ (blue), respectively. Figure 2(e) shows that with the increase of $M$, the $F_E$ calculated by the CFMM converges rapidly to a stable value and the converged value of $F_E$ is consistent with the result of FEM. Figure 2(f) shows that the $F_E$ increases rapidly with the decrease of nanogap size $g$ [4,10,14]. The calculation in Fig. 2(f) by using the CFMM is very efficient because when scanning the nanogap size $g$ or changing the $d$-parameters, only the modal coefficients [ $u_{l,r}^{(l')}$ and $d_{l,r}^{(l')}$ in Eq. (5)] need to be solved by carrying out the scattering-matrix algorithm [see the replacement (44) and relevant statements], while the waveguide eigenmodes [ $\boldsymbol{\psi}_{l,\pm,r}^{(l')}$ in Eq. (5)] need to be solved only once and can be used repeatedly.

Besides, preserving the advantages of the FMM [25] as a modal method, the proposed CFMM under NEBC can solve the full scattering matrix [the $\mathbf{S}_{L+1}$ in Eq. (15)] of the whole structure via a single computation. The different columns of the full scattering matrix describe the excitation of electromagnetic field by different modes, and all columns form a numerically complete set of basis for expressing the electromagnetic field under various excitations. In comparison, for some other full-wave numerical methods such as finite-difference time-domain method [23] or FEM [24], a single computation can only solve the electromagnetic field under a specific excitation, which corresponds to one column or a specific linear combination of some columns of the full scattering matrix (for single- or multi-mode excitation, respectively). Benefiting from the complete description of various excitations, the solution of the full scattering matrix can also realize an efficient and rigorous electromagnetic modeling of some difficult problems, for instance, semi-infinite $z$-periodic structures [64] or finite $z$-periodic structures with many periods [56] and the near-field radiative transfer between two nanostructures [65].

## V. CONCLUSIONS

The FMM and the C method under the NEBC (collectively called the CFMM under NEBC) are proposed by incorporating the NEBC into a recently reported 3D-C method [57] which is applicable to the general 3D photonic structures with curved boundaries. The proposed method can achieve an electromagnetic modeling of the nonclassical quantum effects in the MMNSs while preserving the advantages of the modal methods in physical intuitiveness and computational efficiency compared with other full-wave numerical methods [66]. The validity and accuracy of the proposed method are verified through a comparison with the full-wave FEM incorporating the NEBC [11]. The proposed CFMM under NEBC can be used as an efficient tool for designing plasmonic devices with feature sizes down to extreme nanometer scales [10,11,18,21].

## ACKNOWLEDGEMENT

Financial support from the National Natural Science Foundation of China (NSFC) (62475120, 62075104).

*Corresponding author: liuhht@nankai.edu.cn

# Supplemental Material

# Fourier modal method and coordinate transformation method under nonclassical electromagnetic boundary condition for the electromagnetism of mesoscale metallic nanostructures

Haitao Liu[1,2,*]

[1]*Institute of Modern Optics, College of Electronic Information and Optical Engineering, Nankai University, Tianjin, 300350, China*

[2]*Tianjin Key Laboratory of Micro-scale Optical Information Science and Technology, Tianjin, 300350, China*

[*]Corresponding author: liuht@nankai.edu.cn

The Supplemental Material provides the Fourier representation of the covariant-form Maxwell's equations in the 3D-C method (Section S1), the Fourier representation of the $E_{123}^{\pm}$ defined in Eq. (22) in the main text (Section S2), the Fourier representation of the $D_{23}^{\pm}$ defined in Eq. (28a) in the main text (Section S3), and the Fourier representation of the $D_{13}^{\pm}$ defined in Eq. (28b) in the main text (Section S4).

## S1. FOURIER REPRESENTATION OF THE COVARIANT-FORM MAXWELL'S EQUATIONS IN THE 3D-C METHOD

In this section, we will provide the Fourier representation of the covariant-form Maxwell's equations [Eq. (2) in the main text] in the 3D-C method [1], i.e., the differential equations satisfied by the Fourier expansion coefficients of the covariant components $E_i$ and $H_i$ of electromagnetic field. The main aim of this section is to provide some equations used for deriving the Fourier representation of the tangential electric-field or magnetic-field NEBC under the curvilinear coordinate system of CFMM (in Secs. IIIC and IIID of the main text). Besides, the content of this section is not provided in Ref. [1] (the original reference reporting the 3D-C method under CEBC) for the sake of compactness, and some details of applying the correct Fourier factorization rules [2] are different from Ref. [3].

### A. Continuity of the electromagnetic field components with respect to the curvilinear coordinates $u^1$ and $u^2$ in each $u^3$-invariant layer

Equation (2) in the main text can be rewritten as,

$$\begin{vmatrix} \mathbf{e}_1 & \mathbf{e}_2 & \mathbf{e}_3 \\ \frac{\partial}{\partial u^1} & \frac{\partial}{\partial u^2} & \frac{\partial}{\partial u^3} \\ E_1 & E_2 & E_3 \end{vmatrix} = i\omega(\mathbf{e}_1,\mathbf{e}_2,\mathbf{e}_3)\begin{bmatrix} B_1 \\ B_2 \\ B_3 \end{bmatrix} \text{ (a),} \quad \begin{vmatrix} \mathbf{e}_1 & \mathbf{e}_2 & \mathbf{e}_3 \\ \frac{\partial}{\partial u^1} & \frac{\partial}{\partial u^2} & \frac{\partial}{\partial u^3} \\ H_1 & H_2 & H_3 \end{vmatrix} = -i\omega(\mathbf{e}_1,\mathbf{e}_2,\mathbf{e}_3)\begin{bmatrix} D_1 \\ D_2 \\ D_3 \end{bmatrix} \text{ (b),} \qquad \text{(S1.1)}$$

where we define

$$\begin{bmatrix} B_1 \\ B_2 \\ B_3 \end{bmatrix} = \begin{bmatrix} \mu_{1,1} & \mu_{1,2} & \mu_{1,3} \\ \mu_{2,1} & \mu_{2,2} & \mu_{2,3} \\ \mu_{3,1} & \mu_{3,2} & \mu_{3,3} \end{bmatrix}\begin{bmatrix} H_1 \\ H_2 \\ H_3 \end{bmatrix} \text{ (a),} \quad \begin{bmatrix} D_1 \\ D_2 \\ D_3 \end{bmatrix} = \begin{bmatrix} \varepsilon_{1,1} & \varepsilon_{1,2} & \varepsilon_{1,3} \\ \varepsilon_{2,1} & \varepsilon_{2,2} & \varepsilon_{2,3} \\ \varepsilon_{3,1} & \varepsilon_{3,2} & \varepsilon_{3,3} \end{bmatrix}\begin{bmatrix} E_1 \\ E_2 \\ E_3 \end{bmatrix} \text{ (b),} \qquad \text{(S1.2)}$$

and

$$\begin{bmatrix} \mu_{1,1} & \mu_{1,2} & \mu_{1,3} \\ \mu_{2,1} & \mu_{2,2} & \mu_{2,3} \\ \mu_{3,1} & \mu_{3,2} & \mu_{3,3} \end{bmatrix} = V \begin{bmatrix} \mu^{1,1} & \mu^{1,2} & \mu^{1,3} \\ \mu^{2,1} & \mu^{2,2} & \mu^{2,3} \\ \mu^{3,1} & \mu^{3,2} & \mu^{3,3} \end{bmatrix} \text{ (a)}, \quad \begin{bmatrix} \varepsilon_{1,1} & \varepsilon_{1,2} & \varepsilon_{1,3} \\ \varepsilon_{2,1} & \varepsilon_{2,2} & \varepsilon_{2,3} \\ \varepsilon_{3,1} & \varepsilon_{3,2} & \varepsilon_{3,3} \end{bmatrix} = V \begin{bmatrix} \varepsilon^{1,1} & \varepsilon^{1,2} & \varepsilon^{1,3} \\ \varepsilon^{2,1} & \varepsilon^{2,2} & \varepsilon^{2,3} \\ \varepsilon^{3,1} & \varepsilon^{3,2} & \varepsilon^{3,3} \end{bmatrix} \text{ (b)}. \qquad (S1.3)$$

Equation (S1.2b) is just Eq. (25) in the main text. As mentioned before and after Eq. (25) in the main text, here note that $D_i \neq \mathbf{e}_i \cdot \mathbf{D}$, $B_i \neq \mathbf{e}_i \cdot \mathbf{B}$ (with the definitions $\mathbf{D}=\boldsymbol{\varepsilon}\cdot\mathbf{E}$, $\mathbf{B}=\boldsymbol{\mu}\cdot\mathbf{H}$), $\varepsilon_{i,j} \neq \mathbf{e}_i \cdot \boldsymbol{\varepsilon} \cdot \mathbf{e}_j$ and $\mu_{i,j} \neq \mathbf{e}_i \cdot \boldsymbol{\mu} \cdot \mathbf{e}_j$. There are $D_i = VD^i$ and $B_i = VB^i$ with the definitions $D^i = \mathbf{e}^i \cdot \mathbf{D}$ and $B^i = \mathbf{e}^i \cdot \mathbf{B}$. By virtue of the well-known form invariance of Maxwell's equations [4-6], i.e., the consistency between the form of Eq. (S1.1) along with Eq. (S1.2) under a curvilinear coordinate system $(u^1, u^2, u^3) = (u, v, w)$ and the form of Maxwell's equations under the Cartesian coordinate system $(x_1, x_2, x_3) = (x, y, z)$, one can immediately obtain the following claims concerning the $u^1$- and $u^2$-continuity of electromagnetic field components in each $u^3$-invariant layer where the $\boldsymbol{\varepsilon}$ and $\boldsymbol{\mu}$ (and resultantly, $\varepsilon_{i,j}$ and $\mu_{i,j}$) are invariant with respect to $u^3$:

*Claim S1*: The field components $(D_1, E_2, B_1, H_2)$ are continuous with respect to $u^1$ but possibly not to $u^2$;

*Claim S2*: The field components $(E_1, D_2, H_1, B_2)$ are continuous with respect to $u^2$ but possibly not to $u^1$;

*Claim S3*: The field components $(E_3, H_3)$ are continuous with respect to both $u^1$ and $u^2$;

*Claim S4*: The field components $(D_3, B_3)$ are possibly not continuous with respect to either $u^1$ or $u^2$.

Based on the above claims (or more specifically, the claims concerning the field components which are continuous with respect to $u^1$ or $u^2$), the correct Fourier factorization rules [2] can then be applied to obtain the Fourier representation of the Maxwell's equations (S1.1) and (S1.2), as explained in the following subsections B and C.

For the convenience of readers, in the following we provide a brief introduction of conclusions from the correct Fourier factorization rules [2]. Let $f(u')$ denote a periodic function [for instance, the permittivity-related component $\varepsilon_{1;i,j}$ in Eq. (S1.15c)] and $g(u')$ denote a pseudo-periodic function [for instance, the electromagnetic field component $E_i$ in Eq. (S1.15a)], where $f(u')$ and $g(u')$ have the same period. Then for the product $h(u')=f(u')g(u')$ [for instance, the electromagnetic field component $D_i$ in Eq. (S1.15b)], if either $f(u')$ or $g(u')$ is a continuous function, then the Fourier expansion coefficients of $h(u')$ can be given by the Laurent's rule [2],

$$F_{u'}\{h(u')\}_m = \sum_{n=-M_x}^{M_x} F_{u'}\{f(u')\}_{m-n} F_{u'}\{g(u')\}_n, \quad m = -M_x, -M_x+1, \cdots, M_x, \qquad (S1.4)$$

where $M_x$ denotes the truncated Fourier order, and we define an operator $F_{u'}\{\}_m$ of calculating the Fourier expansion coefficient as in Eq. (S1.16). Equation (S1.4) can also be written as an equivalent matrix form,

$$[F_{u'}\{h(u')\}_m]^{m,1} = [F_{u'}\{f(u')\}_{m-n}]^{m,n}[F_{u'}\{g(u')\}_n]^{n,1}, \qquad (S1.5)$$

where we define $[A_{m,n}]^{m,n}$ as a matrix with its $m$-th row and $n$-th column element being $A_{m,n}$, i.e.,

$$([A_{m,n}]^{m,n})_{m,n} = A_{m,n}. \qquad (S1.6)$$

Equation (S1.6) shows that formally, the above defined operator $[\cdot]^{m,n}$ of mapping a sequence $A_{m,n}$ to a matrix $[A_{m,n}]^{m,n}$ and the operator $(\cdot)_{m,n}$ of mapping a matrix $[A_{m,n}]^{m,n}$ to a sequence $A_{m,n}$ are inverse to each other.

If neither $f(u')$ nor $g(u')$ is a continuous function but $h(u')=f(u')g(u')$ is a continuous function, then the Laurent's rule of Eq. (S1.4) is no longer applicable [2]. For this case, we can rewrite the $h(u')=f(u')g(u')$ as the product $g(u')=[1/f(u')]h(u')$ of $1/f(u')$ and $h(u')$, so that the Laurent's rule becomes applicable and gives [using Eq. (S1.5)],

$$[F_{u'}\{g(u')\}_m]^{m,1} = [F_{u'}\{1/f(u')\}_{m-n}]^{m,n}[F_{u'}\{h(u')\}_n]^{n,1}. \qquad (S1.7)$$

Equation (S1.7) forms a set of linear equations with the $F_{u'}\{h(u')\}_n$ as the unknowns, from which we can solve,

$$[F_{u'}\{h(u')\}_m]^{m,1} = \{[F_{u'}\{1/f(u')\}_{m'-n'}]^{m',n'}\}^{-1}[F_{u'}\{g(u')\}_n]^{n,1}, \qquad (S1.8)$$

where a replacement of indices is performed. By applying the operator $(\cdot)_{m,1}$ (taking the $m$-th element of a column vector) to both sides of Eq. (S1.8), we can obtain,

$$F_{u'}\{h(u')\}_m = \sum_{n=-M_x}^{M_x} (\{[F_{u'}\{1/f(u')\}_{m'-n'}]^{m',n'}\}^{-1})_{m,n} F_{u'}\{g(u')\}_n, \quad m = -M_x, -M_x+1, \cdots, M_x. \tag{S1.9}$$

Equation (S1.8) or its equivalent Eq. (S1.9) is just the inverse rule [2]. It is notable that the use of our defined operators $F_{u'}\{\cdot\}_m$, $[\cdot]^{m,n}$ and $(\cdot)_{m,n}$ can provide a concise albeit precise expression for the inverse rule. Equations (S1.7) and (S1.8) show some flexibility for applying the inverse rule, i.e., besides applying Eq. (S1.9) directly, the inverse rule can also be applied simply by first applying the Laurent's rule [Eq. (S1.7)] and then performing an inverse of matrix [Eq. (S1.8)].

It should be noted that the above introduction is aimed to provide a simple guide for applying the correct Fourier factorization rules [2] without emphasizing the mathematical rigorousness [for instance, the limit or convergence of the right side of Eqs. (S1.4) and (S1.9) to the left side as $M_x \to \infty$ requires an additional proof]. One may refer to Ref. [2] and the related references for a rigorous presentation and proof of the correct Fourier factorization rules.

**B. Fourier representation of the constitutive Eq. (S1.2)**

First, we will derive the Fourier representation of Eq. (S1.2), i.e., the Fourier expansion coefficients of $B_i$ and $D_i$ expressed in terms of the Fourier expansion coefficients of $H_i$ and $E_i$ ($i$=1, 2, 3), respectively. For Eq. (S1.2b), by solving ($E_1$, $D_2$, $D_3$) as the unknowns and treating the ($D_1$, $E_2$, $E_3$) as the known quantities, the former can be expressed in terms of the latter as,

$$\begin{bmatrix} E_1 \\ D_2 \\ D_3 \end{bmatrix} = \begin{bmatrix} \varepsilon_{1;1,1} & \varepsilon_{1;1,2} & \varepsilon_{1;1,3} \\ \varepsilon_{1;2,1} & \varepsilon_{1;2,2} & \varepsilon_{1;2,3} \\ \varepsilon_{1;3,1} & \varepsilon_{1;3,2} & \varepsilon_{1;3,3} \end{bmatrix} \begin{bmatrix} D_1 \\ E_2 \\ E_3 \end{bmatrix}, \tag{S1.10}$$

where

$$\begin{aligned}
&\varepsilon_{1;1,1} = \varepsilon_{1,1}^{-1}, \quad \varepsilon_{1;1,2} = -\varepsilon_{1,1}^{-1}\varepsilon_{1,2}, \quad \varepsilon_{1;1,3} = -\varepsilon_{1,1}^{-1}\varepsilon_{1,3}, \\
&\varepsilon_{1;2,1} = \varepsilon_{2,1}\varepsilon_{1,1}^{-1}, \quad \varepsilon_{1;2,2} = \varepsilon_{2,2} - \varepsilon_{2,1}\varepsilon_{1,1}^{-1}\varepsilon_{1,2}, \quad \varepsilon_{1;2,3} = \varepsilon_{2,3} - \varepsilon_{2,1}\varepsilon_{1,1}^{-1}\varepsilon_{1,3}, \\
&\varepsilon_{1;3,1} = \varepsilon_{3,1}\varepsilon_{1,1}^{-1}, \quad \varepsilon_{1;3,2} = \varepsilon_{3,2} - \varepsilon_{3,1}\varepsilon_{1,1}^{-1}\varepsilon_{1,2}, \quad \varepsilon_{1;3,3} = \varepsilon_{3,3} - \varepsilon_{3,1}\varepsilon_{1,1}^{-1}\varepsilon_{1,3}.
\end{aligned} \tag{S1.11}$$

Equation (S1.11) can be rewritten as a concise form of,

$$\boldsymbol{\varepsilon}_1 = l_1(\boldsymbol{\varepsilon}_0), \tag{S1.12}$$

where $\boldsymbol{\varepsilon}_1$ and $\boldsymbol{\varepsilon}_0$ are two matrices with elements $(\boldsymbol{\varepsilon}_1)_{i,j}=\varepsilon_{1;i,j}$ and $(\boldsymbol{\varepsilon}_0)_{i,j}=\varepsilon_{i,j}$, and an operator $l_\tau$ ($\tau$=1, 2, 3) is defined such that $\mathbf{B}=l_\tau(\mathbf{A})$ is given by

$$(\mathbf{B})_{\rho,\sigma} = \begin{cases} (\mathbf{A})_{\tau,\tau}^{-1}, & \text{for } \rho = \tau \wedge \sigma = \tau; \\ -(\mathbf{A})_{\tau,\tau}^{-1}(\mathbf{A})_{\tau,\sigma}, & \text{for } \rho = \tau \wedge \sigma \neq \tau; \\ (\mathbf{A})_{\rho,\tau}(\mathbf{A})_{\tau,\tau}^{-1}, & \text{for } \rho \neq \tau \wedge \sigma = \tau; \\ (\mathbf{A})_{\rho,\sigma} - (\mathbf{A})_{\rho,\tau}(\mathbf{A})_{\tau,\tau}^{-1}(\mathbf{A})_{\tau,\sigma}, & \text{for } \rho \neq \tau \wedge \sigma \neq \tau, \end{cases} \tag{S1.13}$$

where the elements $(\mathbf{A})_{\rho,\sigma}$ and $(\mathbf{B})_{\rho,\sigma}$ ($\rho$, $\sigma$=1, 2, 3) of matrices $\mathbf{A}$ and $\mathbf{B}$ can be either a scalar or a matrix. Note that compared with the $l_\tau^\pm$ in Ref. [3] [see Eq. (13) therein], the present $l_\tau$ has a slightly different definition, which has a merit of reducing the number of used operators. As exemplified by Eqs. (S1.2b) and (S1.10), the function of the operator $l_\tau$ is to generate the coefficient matrix $\mathbf{B}=l_\tau(\mathbf{A})$ from the coefficient matrix $\mathbf{A}$, where an exchange of the $\tau$-th elements of the two column vectors related by $\mathbf{A}$ gives the two column vectors related by $\mathbf{B}$.

According to *Claims S1 and S3*, the field components ($D_1$, $E_2$, $E_3$) on the right side of Eq. (S1.10) are all continuous with respect to $u^1$ [and resultantly, to $u'$ because $u^1=G_1(u')$ is a continuous function], so the Laurent's rule [2] can be applied to Eq. (S1.10) to obtain the Fourier expansion coefficients with respect to $u'$ for ($E_1$, $D_2$, $D_3$) in terms of those for ($D_1$, $E_2$, $E_3$),

$$\begin{bmatrix} \bar{\mathbf{E}}_1 \\ \bar{\mathbf{D}}_2 \\ \bar{\mathbf{D}}_3 \end{bmatrix} = \begin{bmatrix} \boldsymbol{\varepsilon}_{2;1,1} & \boldsymbol{\varepsilon}_{2;1,2} & \boldsymbol{\varepsilon}_{2;1,3} \\ \boldsymbol{\varepsilon}_{2;2,1} & \boldsymbol{\varepsilon}_{2;2,2} & \boldsymbol{\varepsilon}_{2;2,3} \\ \boldsymbol{\varepsilon}_{2;3,1} & \boldsymbol{\varepsilon}_{2;3,2} & \boldsymbol{\varepsilon}_{2;3,3} \end{bmatrix} \begin{bmatrix} \bar{\mathbf{D}}_1 \\ \bar{\mathbf{E}}_2 \\ \bar{\mathbf{E}}_3 \end{bmatrix}, \tag{S1.14}$$

where we define column vectors $\bar{\mathbf{E}}_i$, $\bar{\mathbf{D}}_i$ and Toeplitz matrix $\boldsymbol{\varepsilon}_{2;i,j}$ ($i, j$=1, 2, 3) with elements ($m, n$=−$M_x$, −$M_x$+1, …, $M_x$, with $M_x$ being the truncated Fourier order),

$$(\bar{\mathbf{E}}_i)_{m,1} = F_{u'}\{E_i\}_m \text{ (a)}, \quad (\bar{\mathbf{D}}_i)_{m,1} = F_{u'}\{D_i\}_m \text{ (b)}, \quad (\boldsymbol{\varepsilon}_{2;i,j})_{m,n} = F_{u'}\{\varepsilon_{1;i,j}\}_{m-n} \text{ (c)}, \tag{S1.15}$$

and we define an operator $F_{u'}\{\cdot\}_m$ of calculating the Fourier expansion coefficient with respect to $u'$,

$$F_{u'}\{E_i\}_m = \frac{1}{\Lambda'_u} \int_{u'_0}^{u'_0+\Lambda'_u} du' E_i(u',v',w) \exp(-ik'_{u,m}u'), \tag{S1.16}$$

where the quantities $\Lambda'_u$, $u'_0$ and $k'_{u,m}$ have been defined following Eq. (7) in the main text. Note that in Eqs. (S1.15a) and (S1.15b), the pseudo-periodic phase-shift constant $k'_{u,0} \neq 0$ is allowed for the Fourier expansion coefficients of the pseudo-periodic functions of electromagnetic field components $E_i$ and $D_i$ (for instance, for a periodic structure excited by an obliquely incident plane wave), while in Eq. (S1.15c), there is always $k'_{u,0}=0$ for the periodic functions of permittivity-related components $\varepsilon_{1;i,j}$. Equation (S1.15c) can be rewritten as a concise form of,

$$\boldsymbol{\varepsilon}_2 = F_1(\boldsymbol{\varepsilon}_1). \tag{S1.17}$$

From Eq. (S1.14), one can obtain,

$$\begin{bmatrix} \bar{\mathbf{D}}_1 \\ \bar{\mathbf{D}}_2 \\ \bar{\mathbf{D}}_3 \end{bmatrix} = \begin{bmatrix} \boldsymbol{\varepsilon}_{3;1,1} & \boldsymbol{\varepsilon}_{3;1,2} & \boldsymbol{\varepsilon}_{3;1,3} \\ \boldsymbol{\varepsilon}_{3;2,1} & \boldsymbol{\varepsilon}_{3;2,2} & \boldsymbol{\varepsilon}_{3;2,3} \\ \boldsymbol{\varepsilon}_{3;3,1} & \boldsymbol{\varepsilon}_{3;3,2} & \boldsymbol{\varepsilon}_{3;3,3} \end{bmatrix} \begin{bmatrix} \bar{\mathbf{E}}_1 \\ \bar{\mathbf{E}}_2 \\ \bar{\mathbf{E}}_3 \end{bmatrix}, \tag{S1.18}$$

where

$$\boldsymbol{\varepsilon}_3 = l_1(\boldsymbol{\varepsilon}_2). \tag{S1.19}$$

Equation (S1.18) can further give,

$$\begin{bmatrix} \bar{\mathbf{D}}_1 \\ \bar{\mathbf{E}}_2 \\ \bar{\mathbf{D}}_3 \end{bmatrix} = \begin{bmatrix} \boldsymbol{\varepsilon}_{4;1,1} & \boldsymbol{\varepsilon}_{4;1,2} & \boldsymbol{\varepsilon}_{4;1,3} \\ \boldsymbol{\varepsilon}_{4;2,1} & \boldsymbol{\varepsilon}_{4;2,2} & \boldsymbol{\varepsilon}_{4;2,3} \\ \boldsymbol{\varepsilon}_{4;3,1} & \boldsymbol{\varepsilon}_{4;3,2} & \boldsymbol{\varepsilon}_{4;3,3} \end{bmatrix} \begin{bmatrix} \bar{\mathbf{E}}_1 \\ \bar{\mathbf{D}}_2 \\ \bar{\mathbf{E}}_3 \end{bmatrix}, \tag{S1.20}$$

where

$$\boldsymbol{\varepsilon}_4 = l_2(\boldsymbol{\varepsilon}_3). \tag{S1.21}$$

According to *Claims S2 and S3*, the field components ($E_1$, $D_2$, $E_3$) are all continuous with respect to $u^2$ [and resultantly, to $v'$ because $u^2=G_2(v')$ is a continuous function], so the elements of the three column vectors $[\bar{\mathbf{E}}_1; \bar{\mathbf{D}}_2; \bar{\mathbf{E}}_3]$ on the right side of Eq. (S1.20) are all continuous with respect to $v'$. Therefore, the Laurent's rule [2] can be applied to Eq. (S1.20) to obtain the Fourier expansion coefficients with respect to $v'$ for the elements of the three column vectors $[\bar{\mathbf{D}}_1; \bar{\mathbf{E}}_2; \bar{\mathbf{D}}_3]$ in terms of those for the elements of $[\bar{\mathbf{E}}_1; \bar{\mathbf{D}}_2; \bar{\mathbf{E}}_3]$,

$$\begin{bmatrix} \tilde{\mathbf{D}}_1 \\ \tilde{\mathbf{E}}_2 \\ \tilde{\mathbf{D}}_3 \end{bmatrix} = \begin{bmatrix} \boldsymbol{\varepsilon}_{5;1,1} & \boldsymbol{\varepsilon}_{5;1,2} & \boldsymbol{\varepsilon}_{5;1,3} \\ \boldsymbol{\varepsilon}_{5;2,1} & \boldsymbol{\varepsilon}_{5;2,2} & \boldsymbol{\varepsilon}_{5;2,3} \\ \boldsymbol{\varepsilon}_{5;3,1} & \boldsymbol{\varepsilon}_{5;3,2} & \boldsymbol{\varepsilon}_{5;3,3} \end{bmatrix} \begin{bmatrix} \tilde{\mathbf{E}}_1 \\ \tilde{\mathbf{D}}_2 \\ \tilde{\mathbf{E}}_3 \end{bmatrix}, \tag{S1.22}$$

where we define column vectors $\tilde{\mathbf{E}}_i$, $\tilde{\mathbf{D}}_i$ and matrix $\boldsymbol{\varepsilon}_{5;i,j}$ ($i, j$=1, 2, 3) with elements ($m, n=-M_x, -M_x+1, \ldots, M_x$; $p, q=-M_y, -M_y+1, \ldots, M_y$, with $M_x$ and $M_y$ being the truncated Fourier orders),

$$(\tilde{\mathbf{E}}_i)_{(m,p),1} = F_{(u',v')}\{E_i\}_{(m,p)} \text{ (a)}, \quad (\tilde{\mathbf{D}}_i)_{(m,p),1} = F_{(u',v')}\{D_i\}_{(m,p)} \text{ (b)}, \quad (\boldsymbol{\varepsilon}_{5;i,j})_{(m,p),(n,q)} = F_{v'}\{(\boldsymbol{\varepsilon}_{4;i,j})_{m,n}\}_{p-q} \text{ (c)}, \tag{S1.23}$$

and we define an operator $F_{v'}\{\cdot\}_p$ of calculating the Fourier expansion coefficient with respect to $v'$,

$$F_{v'}\{h\}_p = \frac{1}{\Lambda'_v} \int_{v'_0}^{v'_0+\Lambda'_v} dv' h(u',v',w) \exp(-ik'_{v,p}v'), \tag{S1.24}$$

where the quantities $\Lambda'_v$, $v'_0$ and $k'_{v,p}$ have been defined following Eq. (7) in the main text. In Eq. (S1.23), $(\cdot)_{(m,p),1}$ means taking the $(m,p)$-th element of a column vector, and $(\cdot)_{(m,p),(n,q)}$ means taking the $(m,p)$-th row and $(n,q)$-th column element of a matrix, which are similar to the operator $(\cdot)_{m,n}$ defined in Eq. (S1.6) except for using a dual-number index for a row or a column. Similar to Eq. (S1.15), $k'_{v,0} \neq 0$ is allowed for the Fourier expansion coefficients of the pseudo-periodic functions of electromagnetic field components $E_i$ and $D_i$ in Eqs. (S1.23a) and (S1.23b), while in Eq. (S1.23c), there is always $k'_{v,0}=0$ for the periodic functions of permittivity-related components $(\boldsymbol{\varepsilon}_{4;i,j})_{m,n}$. For deriving Eq. (S1.22), we have also used

$$F_{v'}\{F_{u'}\{h\}_m\}_p = F_{(u',v')}\{h\}_{(m,p)}, \tag{S1.25}$$

where the right side has been defined in Eq. (7) in the main text. Equation (S1.23c) can be rewritten as a concise form of,

$$\boldsymbol{\varepsilon}_5 = F_2(\boldsymbol{\varepsilon}_4). \tag{S1.26}$$

From Eq. (S1.22), one can obtain,

$$\begin{bmatrix} \tilde{\mathbf{D}}_1 \\ \tilde{\mathbf{D}}_2 \\ \tilde{\mathbf{D}}_3 \end{bmatrix} = \begin{bmatrix} \boldsymbol{\varepsilon}_{6;1,1} & \boldsymbol{\varepsilon}_{6;1,2} & \boldsymbol{\varepsilon}_{6;1,3} \\ \boldsymbol{\varepsilon}_{6;2,1} & \boldsymbol{\varepsilon}_{6;2,2} & \boldsymbol{\varepsilon}_{6;2,3} \\ \boldsymbol{\varepsilon}_{6;3,1} & \boldsymbol{\varepsilon}_{6;3,2} & \boldsymbol{\varepsilon}_{6;3,3} \end{bmatrix} \begin{bmatrix} \tilde{\mathbf{E}}_1 \\ \tilde{\mathbf{E}}_2 \\ \tilde{\mathbf{E}}_3 \end{bmatrix}, \tag{S1.27}$$

where

$$\boldsymbol{\varepsilon}_6 = l_2(\boldsymbol{\varepsilon}_5). \tag{S1.28}$$

Equations (S1.12), (S1.17), (S1.19), (S1.21), (S1.26) and (S1.28) together give,

$$\boldsymbol{\varepsilon}_6 = l_2(F_2(l_2(l_1(F_1(l_1(\boldsymbol{\varepsilon}_0)))))) = (l_2 F_2 l_2 l_1 F_1 l_1)(\boldsymbol{\varepsilon}_0), \tag{S1.29}$$

which is consistent with Eq. (20) in Ref. [3] except for some slight difference (using less types of operators). Equation (S1.27) along with Eq. (S1.29) is the main result of this subsection, i.e., the Fourier representation of the constitutive Eq. (S1.2b). In a fully parallel way (simply by a replacement of notations), one can also obtain the Fourier representation of the constitutive Eq. (S1.2a),

$$\begin{bmatrix} \tilde{\mathbf{B}}_1 \\ \tilde{\mathbf{B}}_2 \\ \tilde{\mathbf{B}}_3 \end{bmatrix} = \begin{bmatrix} \boldsymbol{\mu}_{6;1,1} & \boldsymbol{\mu}_{6;1,2} & \boldsymbol{\mu}_{6;1,3} \\ \boldsymbol{\mu}_{6;2,1} & \boldsymbol{\mu}_{6;2,2} & \boldsymbol{\mu}_{6;2,3} \\ \boldsymbol{\mu}_{6;3,1} & \boldsymbol{\mu}_{6;3,2} & \boldsymbol{\mu}_{6;3,3} \end{bmatrix} \begin{bmatrix} \tilde{\mathbf{H}}_1 \\ \tilde{\mathbf{H}}_2 \\ \tilde{\mathbf{H}}_3 \end{bmatrix}, \tag{S1.30}$$

where we define the column vectors $\tilde{\mathbf{H}}_i$ and $\tilde{\mathbf{B}}_i$ with elements,

$$(\tilde{\mathbf{H}}_i)_{(m,p),1} = F_{(u',v')}\{H_i\}_{(m,p)} \text{ (a)}, \quad (\tilde{\mathbf{B}}_i)_{(m,p),1} = F_{(u',v')}\{B_i\}_{(m,p)} \text{ (b)}, \tag{S1.31}$$

and we define the matrix,

$$\mathbf{\mu}_6 = (l_2 F_2 l_2 l_1 F_1 l_1)(\mathbf{\mu}_0). \tag{S1.32}$$

with $(\mathbf{\mu}_0)_{i,j} = \mu_{i,j}$.

## C. Fourier representation of the covariant-form Maxwell's equations (S1.1)

By applying the operator $F_{(u',v')}\{\cdot\}_{(m,p)}$ of calculating the Fourier expansion coefficient to both sides of Eq. (S1.1) which takes the component of $\mathbf{e}_i$ ($i=1, 2, 3$), one can obtain,

$$ik_0\boldsymbol{\beta}\tilde{\mathbf{E}}_3 - \frac{d\tilde{\mathbf{E}}_2}{dw} = ik_0\tilde{\mathbf{B}}_1 \text{ (a1)}, \quad \frac{d\tilde{\mathbf{E}}_1}{dw} - ik_0\boldsymbol{\alpha}\tilde{\mathbf{E}}_3 = ik_0\tilde{\mathbf{B}}_2 \text{ (a2)}, \quad ik_0\boldsymbol{\alpha}\tilde{\mathbf{E}}_2 - ik_0\boldsymbol{\beta}\tilde{\mathbf{E}}_1 = ik_0\tilde{\mathbf{B}}_3 \text{ (a3)},$$
$$ik_0\boldsymbol{\beta}\tilde{\mathbf{H}}_3 - \frac{d\tilde{\mathbf{H}}_2}{dw} = -ik_0\tilde{\mathbf{D}}_1 \text{ (b1)}, \quad \frac{d\tilde{\mathbf{H}}_1}{dw} - ik_0\boldsymbol{\alpha}\tilde{\mathbf{H}}_3 = -ik_0\tilde{\mathbf{D}}_2 \text{ (b2)}, \quad ik_0\boldsymbol{\alpha}\tilde{\mathbf{H}}_2 - ik_0\boldsymbol{\beta}\tilde{\mathbf{H}}_1 = -ik_0\tilde{\mathbf{D}}_3 \text{ (b3)}, \tag{S1.33}$$

where we define matrices,

$$\boldsymbol{\alpha} = \tilde{\mathbf{g}}_1 \mathbf{K}_1 \text{ (a)}, \quad \boldsymbol{\beta} = \tilde{\mathbf{g}}_2 \mathbf{K}_2 \text{ (b)},$$
$$(\tilde{\mathbf{g}}_1)_{(m,p),(n,q)} = F_{u'}\{g_1(u')\}_{m-n}\delta_{p,q} \text{ (c)}, \quad (\tilde{\mathbf{g}}_2)_{(m,p),(n,q)} = \delta_{m,n}F_{v'}\{g_2(v')\}_{p-q} \text{ (d)}, \tag{S1.34}$$
$$(\mathbf{K}_1)_{(m,p),(n,q)} = (k'_{u,m}/k_0)\delta_{m,n}\delta_{p,q} \text{ (e)}, \quad (\mathbf{K}_2)_{(m,p),(n,q)} = (k'_{v,p}/k_0)\delta_{m,n}\delta_{p,q} \text{ (f)}.$$

In Eq. (S1.33), the $\omega$ in Eq. (S1.1) has been replaced by $k_0$ for the Gaussian system of units used in this paper, and the setting of $g_1(u')$ and $g_2(v')$ as continuous functions has been used to apply the Laurent's rule [2] for the Fourier factorization of the products in the right side of Eq. (4) in the main text. To eliminate the $\tilde{\mathbf{E}}_3$ and $\tilde{\mathbf{H}}_3$ in Eqs. (S1.33a1), (S1.33a2), (S1.33b1) and (S1.33b2), Eq. (S1.27) gives,

$$\begin{bmatrix} \tilde{\mathbf{D}}_1 \\ \tilde{\mathbf{D}}_2 \\ \tilde{\mathbf{E}}_3 \end{bmatrix} = \begin{bmatrix} \boldsymbol{\varepsilon}_{7;1,1} & \boldsymbol{\varepsilon}_{7;1,2} & \boldsymbol{\varepsilon}_{7;1,3} \\ \boldsymbol{\varepsilon}_{7;2,1} & \boldsymbol{\varepsilon}_{7;2,2} & \boldsymbol{\varepsilon}_{7;2,3} \\ \boldsymbol{\varepsilon}_{7;3,1} & \boldsymbol{\varepsilon}_{7;3,2} & \boldsymbol{\varepsilon}_{7;3,3} \end{bmatrix} \begin{bmatrix} \tilde{\mathbf{E}}_1 \\ \tilde{\mathbf{E}}_2 \\ \tilde{\mathbf{D}}_3 \end{bmatrix}, \tag{S1.35}$$

where

$$\boldsymbol{\varepsilon}_7 = l_3(\boldsymbol{\varepsilon}_6), \tag{S1.36}$$

and Eq. (S1.30) gives,

$$\begin{bmatrix} \tilde{\mathbf{B}}_1 \\ \tilde{\mathbf{B}}_2 \\ \tilde{\mathbf{H}}_3 \end{bmatrix} = \begin{bmatrix} \boldsymbol{\mu}_{7;1,1} & \boldsymbol{\mu}_{7;1,2} & \boldsymbol{\mu}_{7;1,3} \\ \boldsymbol{\mu}_{7;2,1} & \boldsymbol{\mu}_{7;2,2} & \boldsymbol{\mu}_{7;2,3} \\ \boldsymbol{\mu}_{7;3,1} & \boldsymbol{\mu}_{7;3,2} & \boldsymbol{\mu}_{7;3,3} \end{bmatrix} \begin{bmatrix} \tilde{\mathbf{H}}_1 \\ \tilde{\mathbf{H}}_2 \\ \tilde{\mathbf{B}}_3 \end{bmatrix}, \tag{S1.37}$$

where

$$\boldsymbol{\mu}_7 = l_3(\boldsymbol{\mu}_6). \tag{S1.38}$$

Substituting Eq. (S1.33b3) into Eq. (S1.35), one can obtain,

$$\begin{bmatrix} \tilde{\mathbf{D}}_1 \\ \tilde{\mathbf{D}}_2 \\ \tilde{\mathbf{E}}_3 \end{bmatrix} = \begin{bmatrix} \boldsymbol{\chi}_{1;1,1} & \boldsymbol{\chi}_{1;1,2} & \boldsymbol{\chi}_{1;1,3} & \boldsymbol{\chi}_{1;1,4} \\ \boldsymbol{\chi}_{1;2,1} & \boldsymbol{\chi}_{1;2,2} & \boldsymbol{\chi}_{1;2,3} & \boldsymbol{\chi}_{1;2,4} \\ \boldsymbol{\chi}_{1;3,1} & \boldsymbol{\chi}_{1;3,2} & \boldsymbol{\chi}_{1;3,3} & \boldsymbol{\chi}_{1;3,4} \end{bmatrix} \begin{bmatrix} \tilde{\mathbf{E}}_1 \\ \tilde{\mathbf{E}}_2 \\ \tilde{\mathbf{H}}_1 \\ \tilde{\mathbf{H}}_2 \end{bmatrix}, \tag{S1.39}$$

where

$$\chi_{1;1,1} = \varepsilon_{7;1,1}, \; \chi_{1;1,2} = \varepsilon_{7;1,2}, \; \chi_{1;1,3} = \varepsilon_{7;1,3}\beta, \; \chi_{1;1,4} = -\varepsilon_{7;1,3}\alpha,$$
$$\chi_{1;2,1} = \varepsilon_{7;2,1}, \; \chi_{1;2,2} = \varepsilon_{7;2,2}, \; \chi_{1;2,3} = \varepsilon_{7;2,3}\beta, \; \chi_{1;2,4} = -\varepsilon_{7;2,3}\alpha, \quad (S1.40)$$
$$\chi_{1;3,1} = \varepsilon_{7;3,1}, \; \chi_{1;3,2} = \varepsilon_{7;3,2}, \; \chi_{1;3,3} = \varepsilon_{7;3,3}\beta, \; \chi_{1;3,4} = -\varepsilon_{7;3,3}\alpha.$$

Substituting Eq. (S1.33a3) into Eq. (S1.37), one can obtain,

$$\begin{bmatrix} \tilde{B}_1 \\ \tilde{B}_2 \\ \tilde{H}_3 \end{bmatrix} = \begin{bmatrix} \chi_{1;4,1} & \chi_{1;4,2} & \chi_{1;4,3} & \chi_{1;4,4} \\ \chi_{1;5,1} & \chi_{1;5,2} & \chi_{1;5,3} & \chi_{1;5,4} \\ \chi_{1;6,1} & \chi_{1;6,2} & \chi_{1;6,3} & \chi_{1;6,4} \end{bmatrix} \begin{bmatrix} \tilde{E}_1 \\ \tilde{E}_2 \\ \tilde{H}_1 \\ \tilde{H}_2 \end{bmatrix}, \quad (S1.41)$$

where

$$\chi_{1;4,1} = -\mu_{7;1,3}\beta, \; \chi_{1;4,2} = \mu_{7;1,3}\alpha, \; \chi_{1;4,3} = \mu_{7;1,1}, \; \chi_{1;4,4} = \mu_{7;1,2},$$
$$\chi_{1;5,1} = -\mu_{7;2,3}\beta, \; \chi_{1;5,2} = \mu_{7;2,3}\alpha, \; \chi_{1;5,3} = \mu_{7;2,1}, \; \chi_{1;5,4} = \mu_{7;2,2}, \quad (S1.42)$$
$$\chi_{1;6,1} = -\mu_{7;3,3}\beta, \; \chi_{1;6,2} = \mu_{7;3,3}\alpha, \; \chi_{1;6,3} = \mu_{7;3,1}, \; \chi_{1;6,4} = \mu_{7;3,2}.$$

Substituting Eqs. (S1.39) and (S1.41) into Eqs. (S1.33a1), (S1.33a2), (S1.33b1) and (S1.33b2), one can obtain,

$$\frac{d}{dw'} \begin{bmatrix} \tilde{E}_1 \\ \tilde{E}_2 \\ \tilde{H}_1 \\ \tilde{H}_2 \end{bmatrix} = \begin{bmatrix} \chi_{2;1,1} & \chi_{2;1,2} & \chi_{2;1,3} & \chi_{2;1,4} \\ \chi_{2;2,1} & \chi_{2;2,2} & \chi_{2;2,3} & \chi_{2;2,4} \\ \chi_{2;3,1} & \chi_{2;3,2} & \chi_{2;3,3} & \chi_{2;3,4} \\ \chi_{2;4,1} & \chi_{2;4,2} & \chi_{2;4,3} & \chi_{2;4,4} \end{bmatrix} \begin{bmatrix} \tilde{E}_1 \\ \tilde{E}_2 \\ \tilde{H}_1 \\ \tilde{H}_2 \end{bmatrix} \quad (S1.43)$$

where $w' = k_0 w$ and

$$\chi_{2;1,1} = i\chi_{1;5,1} + i\alpha\chi_{1;3,1}, \; \chi_{2;1,2} = i\chi_{1;5,2} + i\alpha\chi_{1;3,2}, \; \chi_{2;1,3} = i\chi_{1;5,3} + i\alpha\chi_{1;3,3}, \; \chi_{2;1,4} = i\chi_{1;5,4} + i\alpha\chi_{1;3,4},$$
$$\chi_{2;2,1} = -i\chi_{1;4,1} + i\beta\chi_{1;3,1}, \; \chi_{2;2,2} = -i\chi_{1;4,2} + i\beta\chi_{1;3,2}, \; \chi_{2;2,3} = -i\chi_{1;4,3} + i\beta\chi_{1;3,3}, \; \chi_{2;2,4} = -i\chi_{1;4,4} + i\beta\chi_{1;3,4}, \quad (S1.44)$$
$$\chi_{2;3,1} = -i\chi_{1;2,1} + i\alpha\chi_{1;6,1}, \; \chi_{2;3,2} = -i\chi_{1;2,2} + i\alpha\chi_{1;6,2}, \; \chi_{2;3,3} = -i\chi_{1;2,3} + i\alpha\chi_{1;6,3}, \; \chi_{2;3,4} = -i\chi_{1;2,4} + i\alpha\chi_{1;6,4},$$
$$\chi_{2;4,1} = i\chi_{1;1,1} + i\beta\chi_{1;6,1}, \; \chi_{2;4,2} = i\chi_{1;1,2} + i\beta\chi_{1;6,2}, \; \chi_{2;4,3} = i\chi_{1;1,3} + i\beta\chi_{1;6,3}, \; \chi_{2;4,4} = i\chi_{1;1,4} + i\beta\chi_{1;6,4}.$$

Equation (S1.43) is the Fourier representation of the covariant-form Maxwell's equations [Eq. (2) in the main text] in the 3D-C method [1]. By solving Eq. (S1.43), one can obtain the modal solution of electromagnetic field in each $u^3$-invariant layer as given by Eq. (5) in the main text, where the $ik_{l,\pm,r}^{(l')}/k_0$ and $F_{(u',v')}\{\psi_{l,\pm,r}^{(l')}\}_{(m,p)}$ are respectively given by the eigenvalue and the corresponding eigenvector of the coefficient matrix in Eq. (S1.43).

## S2. FOURIER REPRESENTATION OF THE $E_{123}^{\pm}$ DEFINED IN EQ. (22) IN THE MAIN TEXT

Substituting Eq. (S1.10) for $E_1$ into Eq. (22) in the main text, one can obtain,

$$E_{123} = Q_{1,1}D_1 + Q_{1,2}E_2 + Q_{1,3}E_3, \quad (S2.1)$$

where the superscripts + and – have been neglected for brevity, and

$$Q_{1,1} = d'_\perp \frac{g^{3,1}}{|\mathbf{e}^3|}\varepsilon_{1;1,1} \text{ (a), } Q_{1,2} = d'_\perp \frac{g^{3,1}}{|\mathbf{e}^3|}\varepsilon_{1;1,2} + d'_\perp \frac{g^{3,2}}{|\mathbf{e}^3|} \text{ (b), } Q_{1,3} = d'_\perp \frac{g^{3,1}}{|\mathbf{e}^3|}\varepsilon_{1;1,3} + d'_\perp \frac{g^{3,3}}{|\mathbf{e}^3|} \text{ (c).} \quad (S2.2)$$

Then according to *Claims S1 and S3*, the field components ($D_1$, $E_2$, $E_3$) on the right side of Eq. (S2.1) are all continuous with respect to $u^1$ [and resultantly, to $u'$ because $u^1 = G_1(u')$ is a continuous function], so the Laurent's rule [2] can be applied to Eq. (S2.1) to obtain the Fourier expansion coefficients with respect to $u'$ for $E_{123}$ in terms of those for $(D_1, E_2, E_3)$,

$$\bar{\mathbf{E}}_{123} = \mathbf{Q}_{2,1}\bar{\mathbf{D}}_1 + \mathbf{Q}_{2,2}\bar{\mathbf{E}}_2 + \mathbf{Q}_{2,3}\bar{\mathbf{E}}_3, \tag{S2.3}$$

where the column vectors $\bar{\mathbf{E}}_i$ and $\bar{\mathbf{D}}_i$ ($i=1, 2, 3$) have been defined in Eqs. (S1.15a) and (S1.15b), and the column vector $\bar{\mathbf{E}}_{123}$ and matrices $\mathbf{Q}_{2,i}$ ($i=1, 2, 3$) are defined with elements ($m, n=-M_x, -M_x+1, \ldots, M_x$, with $M_x$ being the truncated Fourier order),

$$(\bar{\mathbf{E}}_{123})_{m,1} = F_{u'}\{E_{123}\}_m \text{ (a)}, \quad (\mathbf{Q}_{2,i})_{m,n} = F_{u'}\{Q_{1,i}\}_{m-n} \text{ (b)}. \tag{S2.4}$$

Substituting Eq. (S1.20) for $\bar{\mathbf{D}}_1$ and $\bar{\mathbf{E}}_2$ into Eq. (S2.3), one can obtain,

$$\bar{\mathbf{E}}_{123} = \mathbf{Q}_{3,1}\bar{\mathbf{E}}_1 + \mathbf{Q}_{3,2}\bar{\mathbf{D}}_2 + \mathbf{Q}_{3,3}\bar{\mathbf{E}}_3, \tag{S2.5}$$

where the matrices $\mathbf{Q}_{3,i}$ ($i=1, 2, 3$) are given by,

$$\mathbf{Q}_{3,1} = \mathbf{Q}_{2,1}\boldsymbol{\varepsilon}_{4;1,1} + \mathbf{Q}_{2,2}\boldsymbol{\varepsilon}_{4;2,1} \text{ (a)}, \mathbf{Q}_{3,2} = \mathbf{Q}_{2,1}\boldsymbol{\varepsilon}_{4;1,2} + \mathbf{Q}_{2,2}\boldsymbol{\varepsilon}_{4;2,2} \text{ (b)}, \mathbf{Q}_{3,3} = \mathbf{Q}_{2,1}\boldsymbol{\varepsilon}_{4;1,3} + \mathbf{Q}_{2,2}\boldsymbol{\varepsilon}_{4;2,3} + \mathbf{Q}_{2,3} \text{ (c)}. \tag{S2.6}$$

According to *Claims S2 and S3*, the field components ($E_1, D_2, E_3$) are all continuous with respect to $u^2$ [and resultantly, to $v'$ because $u^2=G_2(v')$ is a continuous function], so the elements of the three column vectors $[\bar{\mathbf{E}}_1; \bar{\mathbf{D}}_2; \bar{\mathbf{E}}_3]$ in the right side of Eq. (S2.5) are all continuous with respect to $v'$. Therefore, the Laurent's rule [2] can be applied to Eq. (S2.5) to obtain the Fourier expansion coefficients with respect to $v'$ for the elements of the column vector $\bar{\mathbf{E}}_{123}$ in terms of those for the elements of $[\bar{\mathbf{E}}_1; \bar{\mathbf{D}}_2; \bar{\mathbf{E}}_3]$,

$$\tilde{\mathbf{E}}_{123} = \mathbf{Q}_{4,1}\tilde{\mathbf{E}}_1 + \mathbf{Q}_{4,2}\tilde{\mathbf{D}}_2 + \mathbf{Q}_{4,3}\tilde{\mathbf{E}}_3, \tag{S2.7}$$

where the column vectors $\tilde{\mathbf{E}}_i$ and $\tilde{\mathbf{D}}_i$ ($i=1, 2, 3$) have been defined in Eqs. (S1.23a) and (S1.23b), and the column vector $\tilde{\mathbf{E}}_{123}$ and matrices $\mathbf{Q}_{4,i}$ ($i=1, 2, 3$) are defined with elements ($m, n=-M_x, -M_x+1, \ldots, M_x; p, q=-M_y, -M_y+1, \ldots, M_y$, with $M_x$ and $M_y$ being the truncated Fourier orders),

$$(\tilde{\mathbf{E}}_{123})_{(m,p),1} = F_{(u',v')}\{E_{123}\}_{(m,p)} \text{ (a)}, \quad (\mathbf{Q}_{4,i})_{(m,p),(n,q)} = F_{v'}\{(\mathbf{Q}_{3,i})_{m,n}\}_{p-q} \text{ (b)}. \tag{S2.8}$$

Substituting Eq. (S1.39) for $\tilde{\mathbf{D}}_2$ and $\tilde{\mathbf{E}}_3$ into Eq. (S2.7), one can obtain,

$$\tilde{\mathbf{E}}_{123} = \mathbf{Q}_{5,1}\tilde{\mathbf{E}}_1 + \mathbf{Q}_{5,2}\tilde{\mathbf{E}}_2 + \mathbf{Q}_{5,3}\tilde{\mathbf{H}}_1 + \mathbf{Q}_{5,4}\tilde{\mathbf{H}}_2, \tag{S2.9}$$

where the matrices $\mathbf{Q}_{5,i}$ ($i=1, 2, 3, 4$) are given by,

$$\mathbf{Q}_{5,1} = \mathbf{Q}_{4,2}\boldsymbol{\chi}_{1;2,1} + \mathbf{Q}_{4,3}\boldsymbol{\chi}_{1;3,1} + \mathbf{Q}_{4,1} \text{ (a)}, \mathbf{Q}_{5,2} = \mathbf{Q}_{4,2}\boldsymbol{\chi}_{1;2,2} + \mathbf{Q}_{4,3}\boldsymbol{\chi}_{1;3,2} \text{ (b)},$$
$$\mathbf{Q}_{5,3} = \mathbf{Q}_{4,2}\boldsymbol{\chi}_{1;2,3} + \mathbf{Q}_{4,3}\boldsymbol{\chi}_{1;3,3} \text{ (c)}, \mathbf{Q}_{5,4} = \mathbf{Q}_{4,2}\boldsymbol{\chi}_{1;2,4} + \mathbf{Q}_{4,3}\boldsymbol{\chi}_{1;3,4} \text{ (d)}. \tag{S2.10}$$

Equation (S2.9) is just Eq. (31) in the main text, which is the main result of this section, i.e., the Fourier representation of the $E_{123}^{\pm}$ defined in Eq. (22) in the main text.

## S3. FOURIER REPRESENTATION OF THE $D_{23}^{\pm}$ DEFINED IN EQ. (28a) IN THE MAIN TEXT

Substituting Eq. (S1.10) for $D_2$ and $D_3$ into Eq. (28a) in the main text, one can obtain,

$$D_{23} = Q_{8,1} D_1 + Q_{8,2} E_2 + Q_{8,3} E_3, \tag{S3.1}$$

where the superscripts + and – have been neglected for brevity, and

$$Q_{8,1} = -id'_{\parallel} \frac{g^{3,3}}{|\mathbf{e}^3|} \varepsilon_{1;2,1} + id'_{\parallel} \frac{g^{3,2}}{|\mathbf{e}^3|} \varepsilon_{1;3,1}, \tag{S3.2a}$$

$$Q_{8,2} = -id'_{\parallel} \frac{g^{3,3}}{|\mathbf{e}^3|} \varepsilon_{1;2,2} + id'_{\parallel} \frac{g^{3,2}}{|\mathbf{e}^3|} \varepsilon_{1;3,2}, \tag{S3.2b}$$

$$Q_{8,3} = -id'_{\parallel} \frac{g^{3,3}}{|\mathbf{e}^3|} \varepsilon_{1;2,3} + id'_{\parallel} \frac{g^{3,2}}{|\mathbf{e}^3|} \varepsilon_{1;3,3}. \tag{S3.2c}$$

Then fully parallel to Eqs. (S2.3), (S2.5), (S2.7) and (S2.9), one can obtain,

$$\tilde{\mathbf{D}}_{23} = \mathbf{Q}_{12,1} \tilde{\mathbf{E}}_1 + \mathbf{Q}_{12,2} \tilde{\mathbf{E}}_2 + \mathbf{Q}_{12,3} \tilde{\mathbf{H}}_1 + \mathbf{Q}_{12,4} \tilde{\mathbf{H}}_2, \tag{S3.3}$$

where the column vector $\tilde{\mathbf{D}}_{23}$ is defined with elements ($m$, $n=-M_x$, $-M_x+1$, …, $M_x$; $p$, $q=-M_y$, $-M_y+1$, …, $M_y$, with $M_x$ and $M_y$ being the truncated Fourier orders),

$$(\tilde{\mathbf{D}}_{23})_{(m,p),1} = F_{(u',v')}\{D_{23}\}_{(m,p)}, \tag{S3.4}$$

and the definitions of matrices $\mathbf{Q}_{9,i}$, $\mathbf{Q}_{10,i}$, $\mathbf{Q}_{11,i}$ ($i$=1, 2, 3) and $\mathbf{Q}_{12,i}$ ($i$=1, 2, 3, 4) are respectively given by Eqs. (S2.4b), (S2.6), (S2.8b) and (S2.10) with a notation replacement,

$$\mathbf{Q}_{1,i} \to \mathbf{Q}_{8,i}, \ \mathbf{Q}_{2,i} \to \mathbf{Q}_{9,i}, \ \mathbf{Q}_{3,i} \to \mathbf{Q}_{10,i}, \ \mathbf{Q}_{4,i} \to \mathbf{Q}_{11,i}, \ \mathbf{Q}_{5,i} \to \mathbf{Q}_{12,i}. \tag{S3.5}$$

Equation (S3.3) is just Eq. (37a) in the main text, i.e., the Fourier representation of the $D_{23}^{\pm}$ defined in Eq. (28a) in the main text.

## S4. FOURIER REPRESENTATION OF THE $D_{13}^{\pm}$ DEFINED IN EQ. (28b) IN THE MAIN TEXT

Substituting Eq. (S1.10) for $D_3$ into Eq. (28b) in the main text, one can obtain,

$$D_{13} = Q_{14,1} D_1 + Q_{14,2} E_2 + Q_{14,3} E_3, \tag{S4.1}$$

where the superscripts + and – have been neglected for brevity, and

$$Q_{14,1} = id'_{\parallel} \frac{g^{3,3}}{|\mathbf{e}^3|} - id'_{\parallel} \frac{g^{3,1}}{|\mathbf{e}^3|} \varepsilon_{1;3,1} \ (a), \ Q_{14,2} = -id'_{\parallel} \frac{g^{3,1}}{|\mathbf{e}^3|} \varepsilon_{1;3,2} \ (b), \ Q_{14,3} = -id'_{\parallel} \frac{g^{3,1}}{|\mathbf{e}^3|} \varepsilon_{1;3,3} \ (c). \tag{S4.2}$$

Then fully parallel to Eqs. (S2.3), (S2.5), (S2.7) and (S2.9), one can obtain,

$$\tilde{\mathbf{D}}_{13} = \mathbf{Q}_{18,1} \tilde{\mathbf{E}}_1 + \mathbf{Q}_{18,2} \tilde{\mathbf{E}}_2 + \mathbf{Q}_{18,3} \tilde{\mathbf{H}}_1 + \mathbf{Q}_{18,4} \tilde{\mathbf{H}}_2, \tag{S4.3}$$

where we define the column vector $\tilde{\mathbf{D}}_{13}$ with elements ($m$, $n=-M_x$, $-M_x+1$, …, $M_x$; $p$, $q=-M_y$, $-M_y+1$, …, $M_y$, with $M_x$ and $M_y$ being the truncated Fourier orders),

$$(\tilde{\mathbf{D}}_{13})_{(m,p),1} = F_{(u',v')}\{D_{13}\}_{(m,p)}, \qquad (S4.4)$$

and the definitions of matrices $\mathbf{Q}_{15,i}$, $\mathbf{Q}_{16,i}$, $\mathbf{Q}_{17,i}$ ($i=1, 2, 3$) and $\mathbf{Q}_{18,i}$ ($i=1, 2, 3, 4$) are respectively given by Eqs. (S2.4b), (S2.6), (S2.8b) and (S2.10) with a notation replacement,

$$\mathbf{Q}_{1,i} \to \mathbf{Q}_{14,i},\ \mathbf{Q}_{2,i} \to \mathbf{Q}_{15,i},\ \mathbf{Q}_{3,i} \to \mathbf{Q}_{16,i},\ \mathbf{Q}_{4,i} \to \mathbf{Q}_{17,i},\ \mathbf{Q}_{5,i} \to \mathbf{Q}_{18,i}. \qquad (S4.5)$$

Equation (S4.3) is just Eq. (37b) in the main text, i.e., the Fourier representation of the $D_{13}^{\pm}$ defined in Eq. (28b) in the main text.